\documentclass[fleqn,usenatbib]{mnras}
\usepackage{epsfig}
\usepackage{epstopdf}
\usepackage{adjustbox}
\usepackage{graphicx}
\usepackage{color}
\usepackage{amsmath,amssymb}

\title[Modelling remnant and restarted sources observed by LOFAR]{The duty cycle of radio galaxies revealed by LOFAR: remnant and restarted radio source populations in the Lockman Hole}
\author[Shabala et al. ]{Stanislav S. Shabala$^{1,2,3}$\thanks{E-mail: stanislav.shabala@utas.edu.au}, et al.\\
$^{1}$School of Natural Sciences, Private Bag 37, University of Tasmania, Hobart, TAS 7001, Australia,\\
$^{2}$Centre for Astrophysics Research, School of Physics, Astronomy and Mathematics, University of Hertfordshire, College Lane
}
\author[Shabala et al. ]{Stanislav S. Shabala$^{1,2,3}$\thanks{E-mail:
		stanislav.shabala@utas.edu.au}, Nika Jurlin$^{4,5}$, Raffaella Morganti$^{4,5}$, Marisa Brienza$^{6,7}$,
		\newauthor
		Martin J. Hardcastle$^2$, Leith E. H. Godfrey, Martin G. H. Krause$^2$, Ross J. Turner$^1$\\
$^{1}$School of Natural Sciences, Private Bag 37, University of Tasmania, Hobart, TAS 7001, Australia.\\
$^{2}$Centre for Astrophysics Research, School of Physics, Astronomy and Mathematics, University of Hertfordshire, College Lane,\\
Hatfield, Hertfordshire, AL10 9AB, UK.\\
$^{3}$ARC Centre of Excellence for All-Sky Astrophysics in 3 Dimensions (ASTRO 3D).\\
$^{4}$ASTRON, the Netherlands Institute for Radio Astronomy, Postbus 2, NL-7990 AA Dwingeloo, the Netherlands.\\
$^{5}$Kapteyn Astronomical Institute, University of Groningen, Postbus 800, NL-9700 AV Groningen, the Netherlands.\\
$^{6}$Dipartimento di Fisica e Astronomia, Universit\`a di Bologna, Via P. Gobetti 93/2, I-40129, Bologna, Italy.\\
$^{7}$INAF - Istituto di Radio Astronomia, Via P. Gobetti 101, I-40129 Bologna, Italy.\\
}

\begin{document}
	
\date{\today}
\pagerange{\pageref{firstpage}--\pageref{lastpage}} \pubyear{2012}
\maketitle
\label{firstpage}	
	
\begin{abstract}

Feedback from radio jets associated with Active Galactic Nuclei (AGN) plays a profound role in the evolution of galaxies. Kinetic power of these radio jets appears to show temporal variation, but the mechanism(s) responsible for this process are not yet clear. Recently, the LOw Frequency ARray (LOFAR) has uncovered large populations of active, remnant and restarted radio jet populations. By focusing on LOFAR data in the Lockman Hole, in this work we use the Radio AGN in Semi-Analytic Environments (RAiSE) dynamical model to present the first self-consistent modelling analysis of active, remnant and restarted radio source populations. Consistent with other recent work, our models predict that remnant radio lobes fade quickly. Any high ($>10$ percent) observed fraction of remnant and restarted sources therefore requires a dominant population of short-lived jets. We speculate that this could plausibly be provided by feedback-regulated accretion.

\end{abstract}


\section{Introduction}

By imparting large amounts of energy and momentum to their surroundings, radio jets from Active Galactic Nuclei (AGNs) play a crucial role in the evolution of their host galaxies and large-scale environments. They are responsible for driving out large amounts of atomic, molecular and ionised gas \citep[e.g.][]{NesvadbaEA08,DasyraCombes11,MorgantiEA13,EmontsEA14,AlataloEA15,VillarMartinEA17,KakkadEA18}, and can both suppress and trigger star formation \citep{CroftEA06,CrockettEA12,GaiblerEA12,RupkeVeilleux13,DuganEA17,MukherjeeEA18}. On larger scales, shocks driven by the global expansion of the radio source \citep[e.g.][]{WorrallEA12,HardcastleKrause13,HardcastleKrause14} and later buoyant rise of jet-inflated radio bubbles \citep{ChurazovEA01,YangReynolds16} quench catastrophic cooling which would otherwise take place in rapidly cooling galaxy clusters \citep{BohringerEA93,FabianEA03,FormanEA05,MittalEA09}. All cosmological galaxy formation models invoke this ``jet mode'' of feedback to explain the suppression of star formation in massive galaxies at late times \citep{CrotonEA06,BowerEA06,ShabalaAlexander09,VogelsbergerEA14,RaoufEA17,WeinbergerEA18,RaoufEA19,MukherjeeEA19}.

Implicitly assumed in all feedback models are the energies and scales (both spatial and temporal) over which the energy injection takes. In principle, observations of the radio galaxy populations encode this information. Using the methodology of \citet{ShabalaAlexander09}, \citet{RaoufEA17} showed that requiring galaxy formation models to reproduce the observed properties of both galaxy and radio jet populations at low redshift (where observational constraints are strongest) can rule out certain AGN feedback models. Detailed modelling is required to interpret the observed radio source properties: as shown in many analytical and numerical investigations, radio lobe luminosity can evolve by more than an order of magnitude over a jet lifetime \citep[e.g.][see also Section~\ref{sec:RAiSE}]{KaiserEA97,ShabalaGodfrey13,HardcastleKrause13,HardcastleKrause14,TurnerShabala15,GodfreyShabala16,Hardcastle18}; such evolution is strongly environment-dependent \citep[e.g.][]{YatesEA18,KrauseEA19b}. These studies show that the intermittency of radio AGN activity is naturally explained if jets in massive ellipticals and clusters operate as thermostats \citep{BestEA05,ShabalaEA08,PopeEA12} - the rate of jet energy injection (as inferred from dynamical models) appears to balance out the cooling of the hot gas \citep{KaiserBest07,ShabalaEA08}. Most of the energy is supplied by the relatively rare, powerful radio sources \citep{TurnerShabala15,HardcastleEA19} associated with massive galaxies - precisely the objects in which feedback is needed. The picture in which every massive elliptical at the centre of a moderate or strong cooling flow \citep{MittalEA09} goes through a similar duty cycle is also qualitatively consistent with observations of double-double radio sources \citep{SchoenmakersEA00,SaripalliEA05,KonarHardcastle13}, where multiple episodes of radio jet activity are seen in the same radio galaxy.

The details of the AGN intermittency are important for both understanding the mechanisms responsible for jet triggering, and inferring the efficiency with which the jets couple to the ambient gas. For example, rapid re-triggering of jet activity allows later bursts of jet plasma to expand rapidly into channels evacuated by previous jet episodes \citep{KonarHardcastle13,WalgEA13}, changing both lobe morphology and feedback efficiency \citep{YatesEA18}. Empirical constraints on the jet duty cycle are therefore crucial to robust interpretation of feedback mechanisms.

The combination of excellent surface brightness sensitivity (30-100\,$\mu$Jy/beam) and high (6 arcsec) resolution at low ($\sim 150$ MHz) frequencies by the LOw Frequency ARray (LOFAR) have recently revolutionised studies of the radio galaxy duty cycle. Detailed studies of individual objects \citep[e.g. ][]{ShulevskiEA12,OrruEA15,ShulevskiEA15,BrienzaEA16,BrienzaEA18} have been complemented by large surveys \citep{MahonyEA16,HardcastleEA16,WilliamsEA18,ShimwellEA19}. These observations have uncovered large populations of active \citep{MahonyEA16,HardcastleEA19,SabaterEA19,MingoEA19,DabhadeEA19}, remnant \citep{BrienzaEA17,MahatmaEA18} and re-started \citep{MahatmaEA19,JurlinEA20} radio galaxies. Cross matching with multi-wavelength catalogues \citep{WilliamsEA19} has yielded large samples with redshifts and host galaxy information. These samples have, for the first time, begun to tackle in a statistically meaningful way the nature of the relationship between active and quiescent phases of jet activity. \citet{MahatmaEA18} found that the remnant fraction corresponds to $\leq 9$ percent of the total radio galaxy population. \citet{GodfreyEA17} and \citet{BrienzaEA17} found that most of their remnants do {\it not} have ultra-steep spectra, implying that the remnants fade below the detection limit faster than their spectra age. Low fractions of double-double radio galaxies \citep[$\sim 4$ percent;][]{MahatmaEA19} again suggest that the remnant phase may only be detectable for a relatively short time after the jets switch off. Complementing and expanding that work, \citet{JurlinEA20} recently reported a high ($13-15$ percent) fraction of radio sources in a low-frequency selected sample to be candidates for restarted activity; in these objects a compact new core would co-exist with remnant lobes. Jurlin et al.'s definition of candidate restarted sources encompasses the double-doubles studied by \citet{MahatmaEA19}\footnote{Using visual inspection, Jurlin et al. find 5 out of 158 sources (3 percent) to have an extended inner core, and clear remnant lobes, in their LOFAR 150 MHz observations. This fraction is consistent with the 4 percent double-double fraction reported by \citet{MahatmaEA19} using similar LOFAR observations of the outer lobes, and higher resolution VLA observations of the inner lobes, noting that those authors did not have strictly enforced cuts in size and flux density.}, while also including sources with younger (and hence more compact) innermost pairs of jets.

In this paper, we explore the implications of the observed active, remnant and restarted radio source populations by combining detailed radio source dynamical models which comprehensively treat relevant loss processes with complete samples of active, remnant and restarted radio galaxies, such as those recently presented by \citet{JurlinEA20}. We show that such complete samples constrain the (otherwise uncertain) parameters of remnant and restarted progenitors. The key result of our work is that selecting the active, remnant and restarted sources using a consistent approach (i.e. from the same observations) provides strong constraints on the plausible range of parameter space for remnant and restarted progenitors, and generates robust predictions for the expected remnant and restarted fraction.

We briefly describe our models and data in Section~\ref{sec:dynModels}. Section~\ref{sec:progenitorModels} constrains the distributions in physical properties (jet powers and ages) of the active population. In Section~\ref{sec:remnants} we make predictions for the remnant and restarted fraction, and discuss our findings in Section~\ref{sec:discussion}. We conclude in Section~\ref{sec:conclusions}.

%MukherjeeEA19 is from eagle group, not Dipanjan.

\section{Dynamical modelling of radio sources}
\label{sec:dynModels}

\subsection{General considerations}

A major difficulty in interpreting the statistics of observed remnant and restarted radio sources relates to the poorly known properties of their progenitor populations. In this work, we address this issue by using a well-defined sample of ``normal'' active radio sources with host galaxy information (Section~\ref{sec:data}), selected in the same way as the remnant and restarted sources. We employ forward modelling with the {\it Radio AGN in Semi-analytic Environments} (RAiSE) code (Sections~\ref{sec:RAiSE} and \ref{sec:modelParams}) to constrain the distributions of lifetimes and jet kinetic powers of these active radio galaxies (Section~\ref{sec:activeSampleParams}), and then use these constrained models to make predictions for the remnant and restarted populations (Section~\ref{sec:remnants}). LOFAR samples of active, remnant and restarted sources in the Lockman Hole are described in more detail by \citet{BrienzaEA17,JurlinEA20}, and the RAiSE model by \citet{TurnerShabala15,ShabalaEA17,TurnerEA18a,TurnerEA18b,TurnerShabala19}; we refer the interested reader to these papers for further details.

\subsection{Radio AGN in Semi-analytic Environments}
\label{sec:RAiSE}

Starting with the seminal work of \citet{Scheuer74}, analytical models of radio galaxies have been used to describe the expansion of jet-inflated cocoons of synchrotron-emitting plasma, and make predictions for the temporal evolution of size, synchrotron luminosity, and radio continuum spectrum for a given set of jet and environment parameters. The radio lobes expand due to overpressure of the lobes with respect to the ambient medium; in lobed Fanaroff-Riley Type I and II \citep[FR-I/II][]{FanaroffRiley74} sources the jets also provide ram pressure along the jet axis. For both FR-II and FR-I sources, the temporal evolution of cocoon dynamics are solved using conservation equations. The radio luminosity is then determined by assuming a scaling between lobe pressure and magnetic field (see Section~\ref{sec:modelParams} below), and calculating the aged spectra of electrons initially shock accelerated by first-order Fermi processes at either the hotspots (for FR-IIs) or flare points (for FR-Is), accounting for losses due to adiabatic expansion, synchrotron radiation, and Inverse Compton upscattering of Cosmic Microwave Background photons. Contribution to integrated synchrotron emissivity in extended radio sources from cores, jets and hotspots is typically no more than a few percent \citep{MullinEA08}, and is usually ignored in such models.

A well-known challenge in radio source modelling \citep[e.g.][]{KaiserEA97,HardcastleKrause13,YatesEA18,KrauseEA19b} is the sensitivity of observable radio source parameters (such as size and radio luminosity) on the atmosphere into which the jets are expanding. First generations of radio source models used either constant \citep{BegelmanCioffi89} or simple power-law environments \citep{KaiserEA97,HeinzEA98} to describe such atmospheres; these models produced self-similar radio sources which are inconsistent with observations \citep{MullinEA08,HardcastleKrause13}, and had limited use in interpretation of observations. More sophisticated treatment of radio source atmospheres naturally reproduces the observed narrowing of FR-II sources due to a rapidly declining atmosphere at large radii \citep{TurnerEA18b,Hardcastle18}. 

In this work, we employ the RAiSE model \citep{TurnerShabala15,TurnerEA18a,TurnerEA18b}. Unlike previous models, RAiSE uses outputs of galaxy formation models (primarily dark matter halo mass) to quantify jet environments. While X-ray observations \citep[e.g.][]{InesonEA17} provide an excellent probe of jet atmospheres, these  are time-consuming and are not possible for large samples. On the other hand, it has recently been shown \citep{RodmanEA19} that halo masses derived through optical galaxy clustering provides an excellent measure of jet environments. The RAiSE model has been shown to reproduce the observed relationship \citep{LedlowOwen96,Best09} between radio luminosity, morphology, and host galaxy properties \citep{TurnerShabala15}; recover sub-equipartition lobe magnetic fields consistent with independent Inverse Compton measurements \citep{InesonEA17,TurnerEA18b}; and in combination with hydrodynamic simulations, reconcile the observed discrepancy between spectral and dynamical ages in powerful radio galaxies \citep{TurnerEA18a}. RAiSE has subsequently been used to test jet production models \citep{TurnerShabala15}, quantify the observability of low-power jets in poor environments \citep{ShabalaEA17}, model remnant lobes \citep{Turner18}, and determine cosmological parameters from radio source observations \citep{TurnerShabala19}.

\subsection{Model parameters}
\label{sec:modelParams}

The RAiSE model predicts the temporal evolution of size, radio luminosity, and the radio spectrum, for each assumed combination of jet kinetic power and environment. We fix several model parameters, as detailed below. As discussed in Section~\ref{sec:activeSampleParams}, our findings are relatively insensitive to the choice of most model parameters, as these are only used to inform the input distribution of the jet power and lifetime distributions of the progenitor (active) populations; choosing a different set of parameters will change the inferred jet powers and ages of the active sample, but not substantially affect the predicted remnant and restarted fractions.

Our assumed model parameters are as follows. We set the initial axial ratio (length divided by width) of the sources to 2.5, consistent with observations of 3CRR FR-II sources \citep{TurnerEA18b}; and the lower cutoff in Lorentz factor of the electron energy distribution $\gamma_{\rm min}=500$ \citep[e.g.][]{GodfreyEA09}. We adopt a power-law injection index of electrons $s=2.04$, which gives a spectral index of $\alpha=(s-1)/2=0.52$; we find that our adopted injection index produces lobe spectral indices consistent with observed populations of LOFAR active sources \citep{MahonyEA16}. We note that, with the exception of the lobe spectral index, our results below depend very weakly on this parameter.

Below, we use these models to infer jet powers and lifetimes of active sources (Section~\ref{sec:activeSample}), and make predictions for the remnant and restarted source populations (Section~\ref{sec:remnants}).

\subsection{Data}
\label{sec:data}

We combine our models with visually identified samples of extended ($>60$~arcsec) radio sources in the Lockman Hole, described in \citet{BrienzaEA17} and \citet{JurlinEA20}. That work found 158 extended sources, consisting of 117 active, 18 candidate remnant and 23 candidate restarted radio galaxies. \citet{JurlinEA20} also provided robust host galaxy (and hence redshift) identifications for approximately two thirds of the sample, and radio morphologies for all sources. In the present work, we restrict our samples to radio sources with identified hosts and redshifts $z<1$. Our final samples consist of 74 active (mostly straight FR-I and FR-II morphology, with some Wide-Angle Tails), 15 candidate remnant and 21 candidate restarted sources. We refer the interested reader to \citet{JurlinEA20} for further details.

Relevant to our modelling, it is well established that observed radio morphology is correlated with both jet and environment properties: FR-II radio sources are preferentially hosted by lower-mass galaxies \citep{LedlowOwen96,MiraghaeiBest17}, in poorer environments \citep{WingBlanton11,GendreEA13,MiraghaeiBest17,MassaroEA18}, and are dominated by radiating particles \citep{CrostonEA18}. FR-Is, on the other hand, are more likely to be hosted by massive ellipticals in clusters, and have a large contribution from non-radiating particles \citep{CrostonEA18}, consistent with jet entrainment on kpc scales \citep{Bicknell95,LaingBridle02,WykesEA15}.

\begin{figure}
\begin{center}
\includegraphics[clip, width=0.45\textwidth]{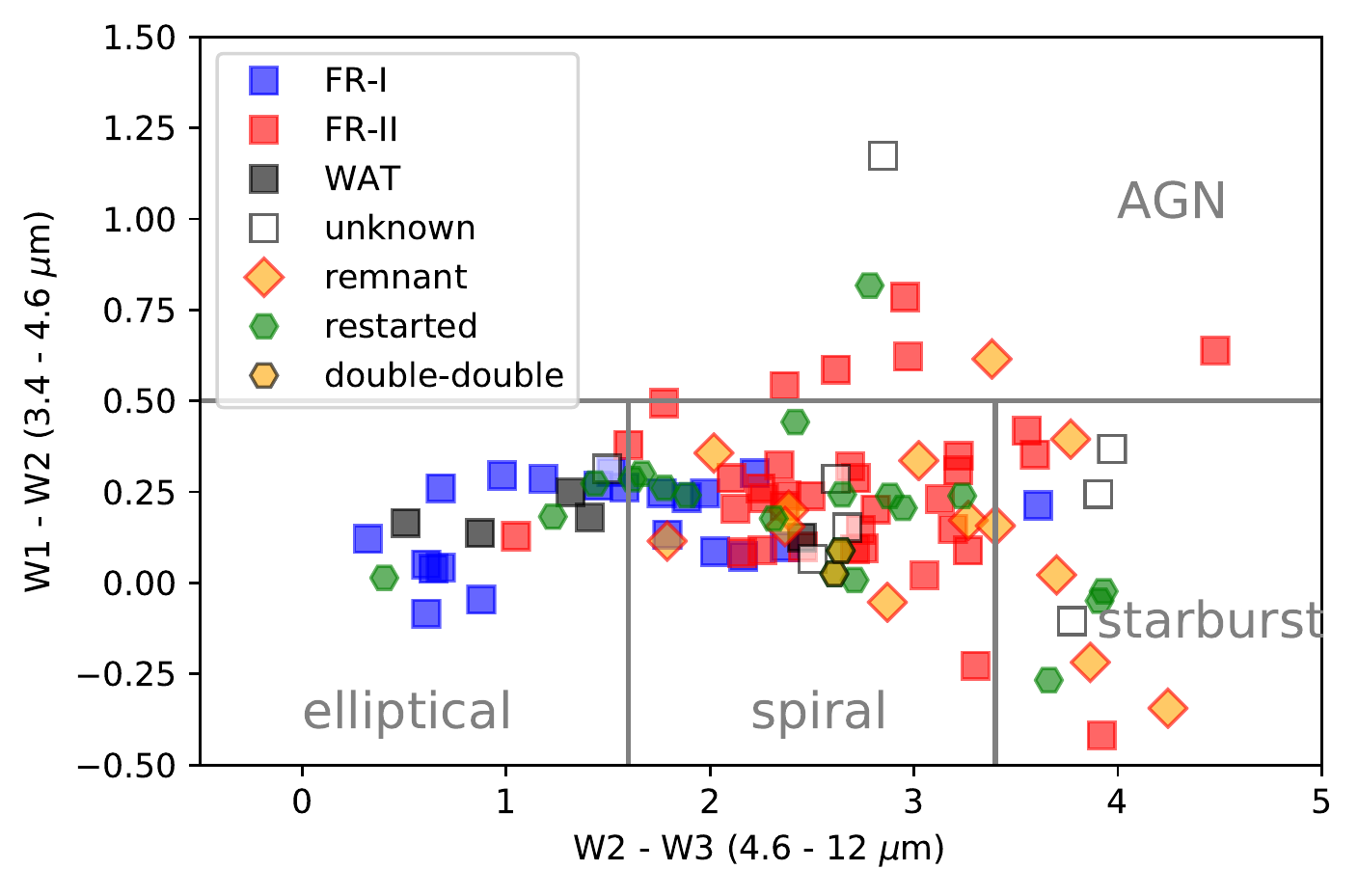}
\caption{WISE colour-colour diagram for the sources considered in this work. 8 of 74 sources with $z<1$ (5 FR-IIs, 2 remnant candidates, 1 restarted candidate) are not detected by WISE, and hence are not plotted here. Classifications follow \citet{MingoEA16}.}
\label{fig:WISE_colour_colour}
\end{center}
\end{figure}

Figure~\ref{fig:WISE_colour_colour} shows the distribution of WISE (W1-W2) vs (W2-W3) colours \citep{WrightEA10} for different radio source morphologies in the Lockman Hole sample. FR-Is and WATs are predominantly (but not exclusively) located in the ``elliptical'' part of the diagram, while FR-IIs are in the ``spiral'' and ``AGN'' parts, consistent with these broadly corresponding to Low- and High-Excitation Radio Galaxy populations, respectively. These results are consistent with those from much larger LOFAR LoTSS \citep{MingoEA19} and Radio Galaxy Zoo \citep{WongEA19} samples. 
These and other \citep[e.g.][]{GurkanEA14} studies have found that radio galaxies accreting in different modes can be explicitly identified using mid-IR diagnostics. Radiatively efficient, High-Excitation Radio Galaxies (HERGs) tend to have bluer, more star-forming hosts, with contributions from hot dust yielding WISE colours in either the ``spiral'' or ``(radiatively efficient) AGN'' parts of the diagram. By contrast, radiatively inefficient Low Excitation Radio Galaxies have very weak mid-IR emission, consistent with a lack of obscuring structure in these objects, placing them primarily in the ``elliptical'' part of the diagram.
Large samples (see e.g. \citet{Tadhunter16} for a review) have confirmed that FR-Is are almost exclusively associated with Low-Excitation host galaxies, while FR-IIs can be hosted by either High or Low-Excitation galaxies. As there is only one FR-II in the ``elliptical'' part of Figure~\ref{fig:WISE_colour_colour}, we treat both the jet and environment properties of FR-I and FR-II populations separately in the remainder of this paper.

% Should WATs be FR-II jets in cluster-like environments?

\section{Constraining models with remnant / restarted progenitors}
\label{sec:progenitorModels}

\subsection{Model tracks}
\label{sec:modelTracks}

To infer the physical properties of the 74 active sources in our sample, we use the RAiSE dynamical model (Section~\ref{sec:RAiSE}) to describe the expansion of jet-inflated lobes. In our models, both FR-II and FR-I radio sources can produce detectable remnants, and we run two separate sets of models for these cases. Consistent with the results of \citet{TurnerEA18b}, we model both sets of sources as jets of pair plasma, with slightly sub-equipartition magnetic fields ($B=0.3 B_{\rm eq}$). We take a galaxy group (halo mass of $10^{13} M_\odot$) as a typical FR-II environment, and distribute the gas according to the self-similar double-beta profile reported by \citet{VikhlininEA06}. FR-II jets are modelled as a standard pair plasma, with non-radiating to radiating particle energy ratio $k \equiv u_p / u_e = 1$. By contrast, FR-Is are modelled as proton dominated with $k \equiv u_p / u_e = 10$, consistent with the median value found by \citet{CrostonEA18}; a $10^{14} M_\odot$ halo is used to represent an FR-I environment. Our choice of environments is guided by the results of \citet{GendreEA13}, who find that FR-Is are preferentially located in clusters and rich groups, while FR-IIs mostly inhabit groups; indicative halo masses are adopted from X-ray observations of \citet{VikhlininEA06,OSullivanEA17}. More realistic assumptions about jet environments would draw from the cluster mass function \citep{Hardcastle18}; however as we show below our predictions for the remnant and restarted source populations are robust to model assumptions, due to our use of the active source population as a constraint.

The effects of different environments and particle content of the FR-II and FR-I jets in our models on observables (e.g. size and luminosity) are not clear: for the same jet kinetic power, the higher gas pressures in the FR-I's cluster environment will result in smaller lobes \citep[e.g.][]{RodmanEA19} and higher luminosities \citep[``environmental boosting",][]{ArnaudEA10,HardcastleKrause13,YatesEA18}, but this will be at least partially compensated by the large fraction of non-radiating particles in these FR-I sources. In Section~\ref{sec:activeSample} we show that these two models make very similar population predictions, and hence our simplified treatment of environments and jet properties is sufficient. We do not model tailed FR-Is, which exhibit quite different dynamics to the lobed FR-II and FR-I populations \citep{LaingBridle02,WangEA09,LaingBridle14,TurnerEA18a}. Tailed FR-Is are typically disrupted on scales of a few kpc \citep{LaingBridle14}; beyond this point their surface brightness sensitivity decreases rapidly \citep{TurnerEA18a}, and hence at the typical redshifts considered here (see Figure~\ref{fig:pred_active_dist_best}) the detectable source sizes will not be large enough to make it into the samples considered in this work.

\begin{figure*}
\begin{center}
\includegraphics[width=0.45\textwidth]{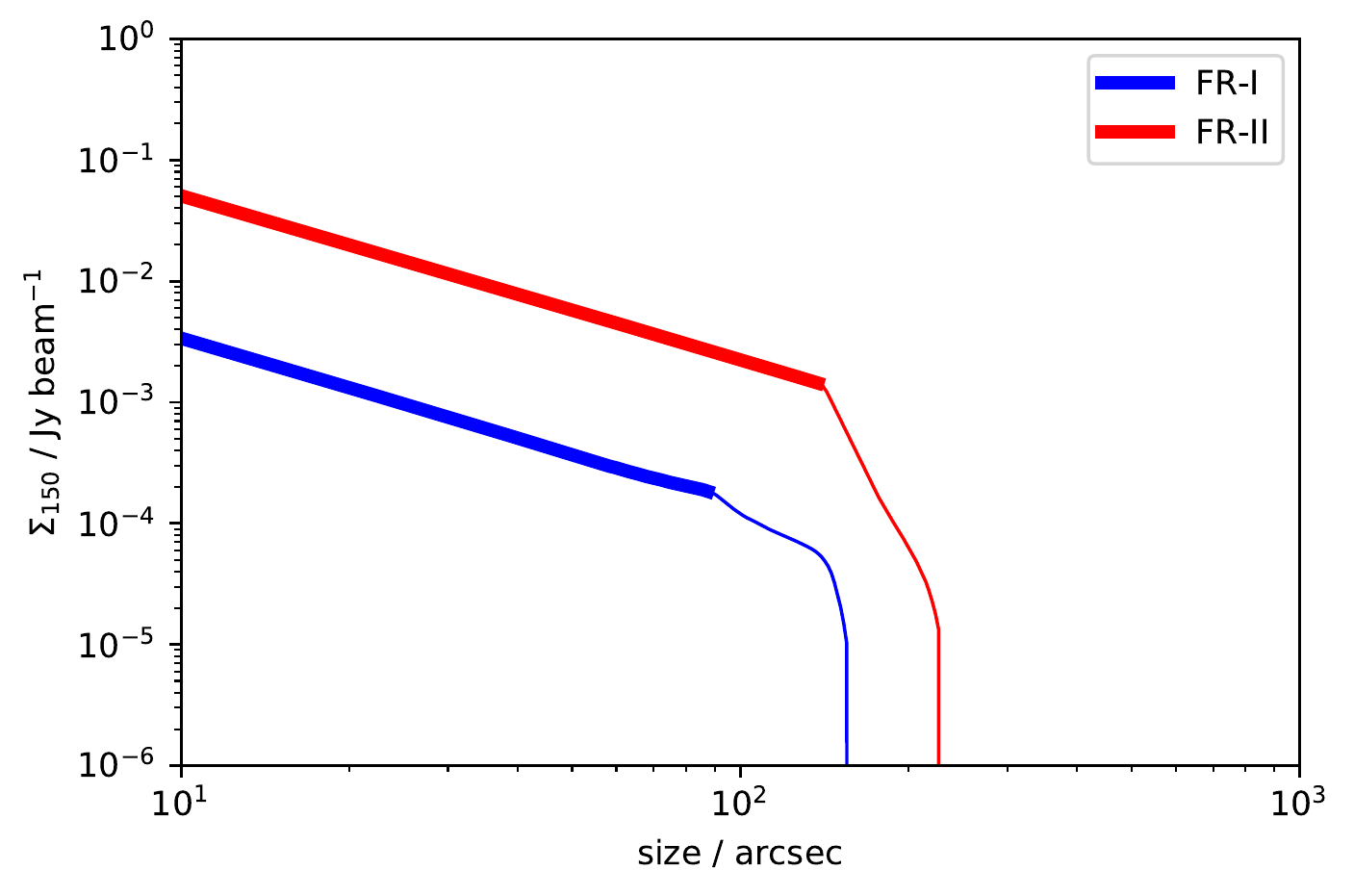}
\includegraphics[width=0.45\textwidth]{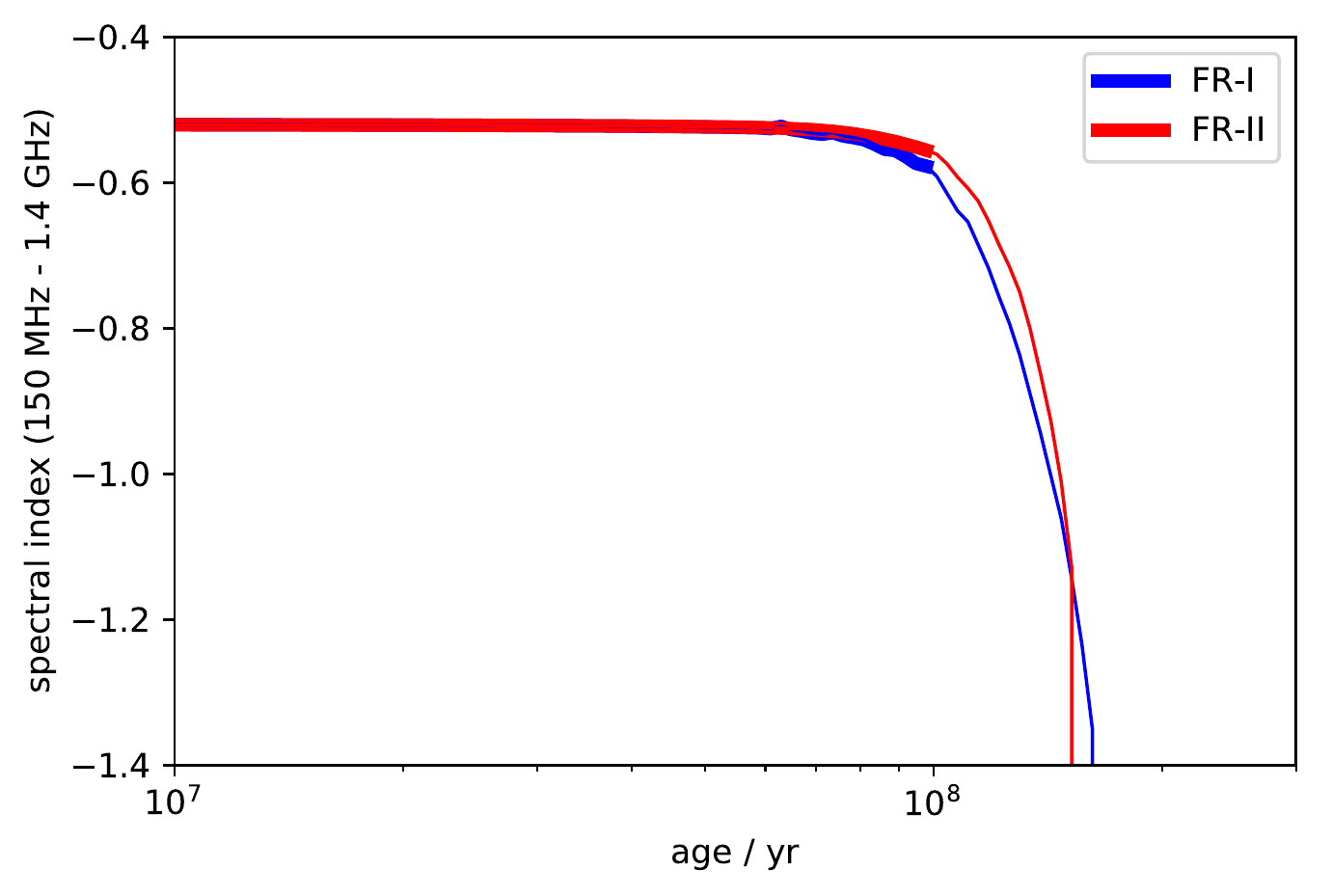}
\caption{Example model tracks at $z=0.2$ for an FR-I jet with $10^{37}$~W kinetic power in a $10^{14} M_\odot$ halo, and an FR-II jet with $10^{38}$~W kinetic power in a $10^{13} M_\odot$ halo. Both jets are on for 100 Myr (thick lines), and evolve as remnants (thin lines) after that time. The lobe surface brightness fades rapidly (left panel), and the lobe spectra steepen (right panel).}
\label{fig:example_tracks}
\end{center}
\end{figure*}

Figure~\ref{fig:example_tracks} shows some representative evolutionary tracks for our models. Lobe luminosity and surface brightness decrease rapidly once the jet is switched off (at 100 Myr in Figure~\ref{fig:example_tracks}), particularly in the case of the more powerful FR-II jet. The spectra, which begin steepening around 50 Myr while the jets are still active, steepen very quickly once the jet is switched off in both models; this behaviour is consistent with previous work \citep[e.g.][]{KaiserCotter02,Turner18,EnglishEA19}.

\subsection{Physical properties of the active sample}
\label{sec:activeSampleParams}

\begin{figure*}
\begin{center}
\includegraphics[width=0.45\textwidth]{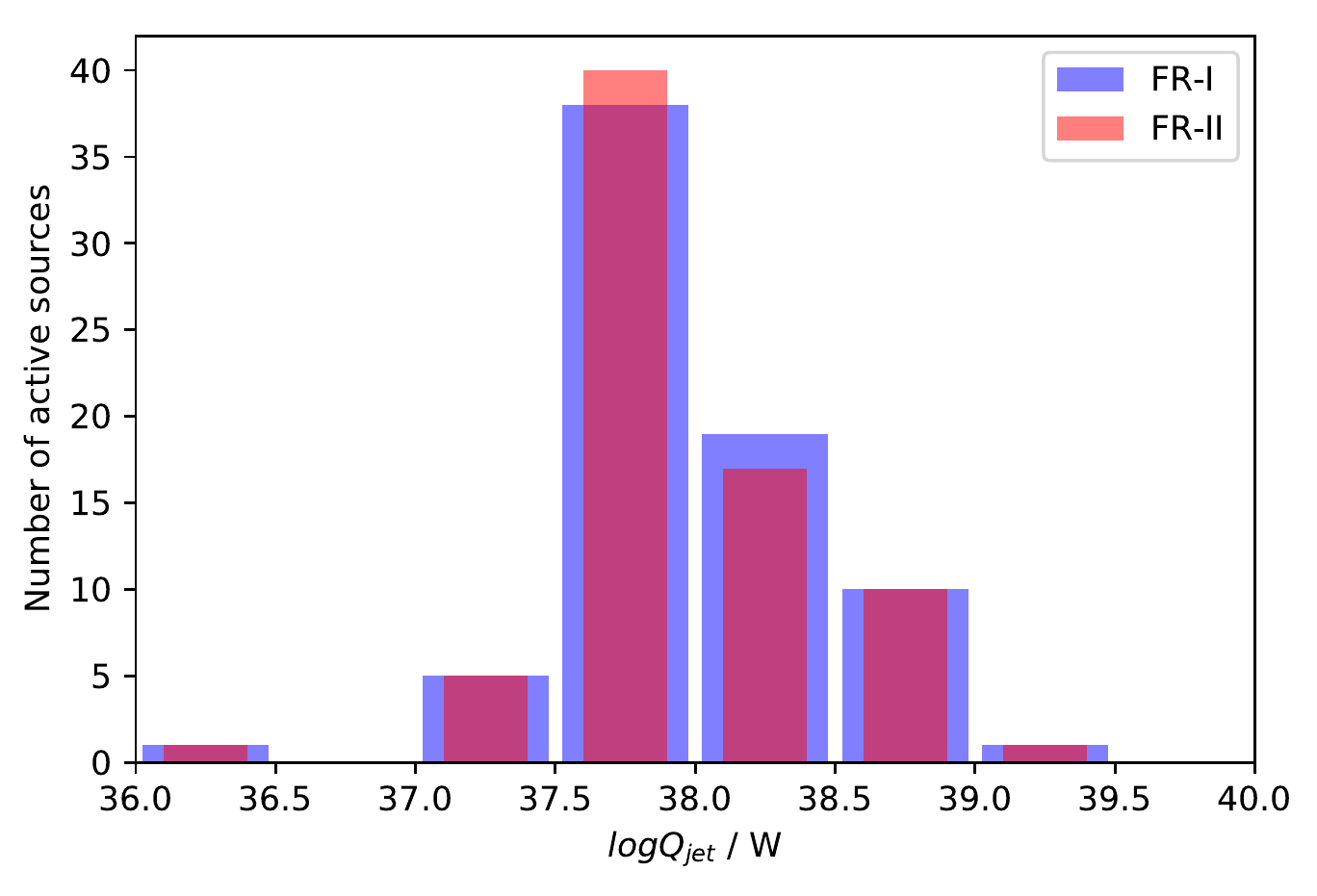}
\includegraphics[width=0.45\textwidth]{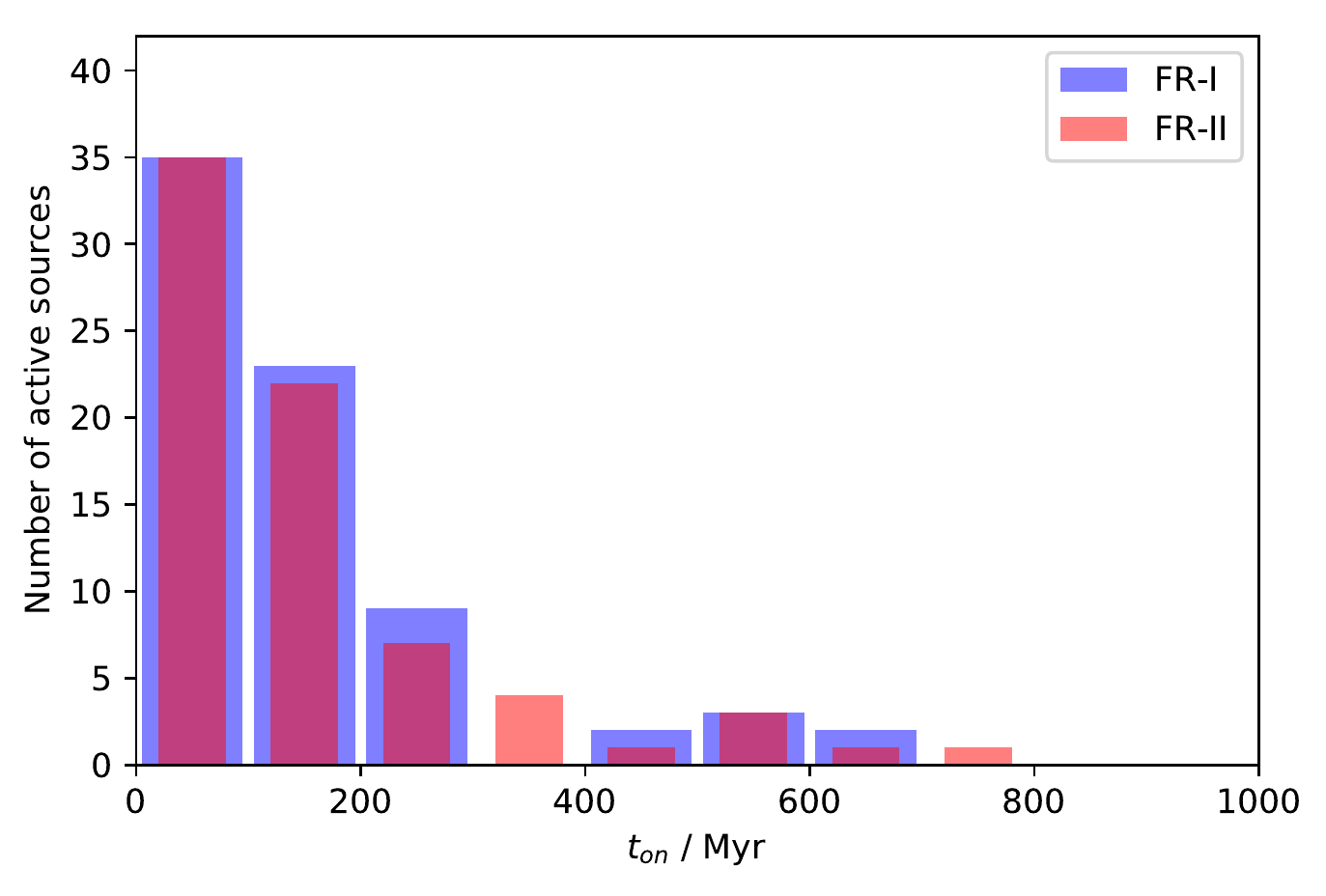}
\caption{Jet kinetic powers and ages for the 74 active sources derived using FR-I and FR-II models for the jets and their environments. All observed sources would be detectable even if they were much older, and hence we are not undersampling the true source age population. Conversely, Malmquist bias results in an apparent correlation between jet power and redshift, and produces an unphysical peak in the inferred jet power distribution. Many more low-power sources are therefore likely to lie below the detection limit.}
\label{fig:samples_physical}
\end{center}
\end{figure*}

Following the approach of \citet{TurnerShabala15}, a grid of RAiSE models is run for each source to a maximum age of 10 Gyr, and a chi-squared minimization procedure is used to recover the best-fitting intrinsic source properties, namely age and jet kinetic power. Our model grids cover jet powers in the range $Q = 10^{35}-10^{40}$~W with spacing $\Delta \log Q = 0.1$ dex; redshift range $z=0.02-1.00$ in steps $\Delta z=0.02$, and ages in the range $t = 10^{3}-10^{10}$~years, with 512 time steps uniformly spaced in $\log t$; this corresponds to a 3 percent age difference between adjacent time steps. The derived jet properties are shown in Figure~\ref{fig:samples_physical}. The apparent peak in jet power distribution is a selection effect: weaker jets are undetectable at large distances over the bulk of their lifetime, and convolving a power-law distribution in jet power \citep[e.g.][]{BrienzaEA17} with a radio detection limit naturally results in such a peaked distribution, as median detectable jet power increases with redshift. On the other hand, the dearth of old ($>400$ Myr) sources is likely to be real: the observed jets are powerful enough to be visible to LOFAR for substantially longer than this time (typically by a factor 2-5), and hence the absence of a population of large, low-surface brightness lobes suggests the rarity of very old sources. We return to this point when discussing the remnant and restarted fraction of sources in Section~\ref{sec:powerLawAgeModels}.

We use the parameters derived in Figure~\ref{fig:samples_physical} to guide the forward modelling in the following sections.

\section{Active radio galaxy population}
\label{sec:activeSample}

\subsection{Distribution in input parameters}
\label{sec:parameterStudy}

In this section, we use forward modelling of radio source populations to constrain the intrinsic physical properties (i.e. jet powers and ages) of the active radio source populations in our Lockman Hole sample, and then use these as inputs to remnant and restarted source modelling.

{\it Jet power}

Following the approach of \citet{BrienzaEA17}, and motivated by the observed decrease in the number of high-power sources (Figure~\ref{fig:samples_physical}), we assume a power-law distribution in the logarithm of jet power, ${\rm prob} (\log Q) \, d \log Q \propto Q^{-a}$. To first order, jet power is correlated with radio luminosity \citep[but see e.g.][]{ShabalaGodfrey13,HardcastleKrause13}, and hence the slope of the AGN Radio Luminosity Function (RLF) allows an estimate of $a$ to be made. \citet{KaiserBest07} used this approach to infer $a \sim 0.6$ for the low-luminosity slope of the RLF; this ignores selection effects against low-power sources, and hence the real distribution is likely to be steeper. Most recently, \citet{HardcastleEA19} found that $a = 1.0$ reproduces well the observed statistics of all but the most luminous radio AGN. Below, we explore a broad range of values $a=0.2 - 1.4$.

{\it Source age}

For source age distributions, we adopt two models. In our first model (Section~\ref{sec:constantAgeModels}), we assume that all sources live to a constant age $t_{\rm on}$. In their analysis of the LOFAR HETDEX field, \citet{HardcastleEA19} employed forward dynamical modeling to infer a median age $t_{\rm on} \sim 500$~Myr for the bright end of their radio AGN sample.
%\footnote{Care must be taken when comparing lifetimes inferred by our models and those of \citet{HardcastleEA19}. On average, the \citet{HardcastleEA19} models sample denser environments, and hence will infer lower jet powers and older ages than this work for a source of given size and lobe luminosity.}.
Below, we explore models for three values of $t_{\rm on} = 0.1, 0.3$ and $1$~Gyr, covering the range of observed ages for active sources (Figure~\ref{fig:samples_physical}).

In our second model (Section~\ref{sec:powerLawAgeModels}), we assume a power-law distribution in age. This is motivated by high observed fractions of compact (on arcsecond scales), low-luminosity sources \citep[e.g.][]{ShabalaEA08,HardcastleEA19}. Allowed ages in our models are as above, but we note that there are implicit cutoffs imposed by our sample selection function: very young sources will be too compact to satisfy the $>60$~arcsec observational cut, while lobe surface brightness will be too low for very old sources to make it into our sample; a similar implicit constraint applies to low jet powers. In Section~\ref{sec:powerLawAgeModels} we show that complete samples of active, remnant and restarted sources can potentially provide excellent constraints on the age distribution function.

{\it Redshift evolution}

Finally, we assume no cosmological evolution in radio source populations across the redshift range of interest ($z=0.2 - 0.9$). This is likely to be a reasonable assumption for at least the low-excitation population \citep{PracyEA16}, and we do not expect this to be a major limitation even for High-Excitation sources given the median redshift of our sample is ``only'' $z \sim 0.5$. In the absence of selection effects, the number of sources detected in a given redshift slice $\Delta z$ should increase with volume as $D_L(z)^2 (D_L(z+\Delta z) - D_L(z)) (1+z)^{-4}$. For concordance cosmology, this corresponds to a flattening in the number counts at $z \sim 0.4-0.5$, followed by approximately constant counts between $z \sim 0.5-1$. Within the assumptions, any turnover in the redshift distribution, as seen in Figures~\ref{fig:pred_active_dist_best} and \ref{fig:pred_active_dist_best_plAge}, is a manifestation of selection effects.

\subsection{Constant age models}
\label{sec:constantAgeModels}

\begin{figure*}
\begin{center}
\includegraphics[width=0.4\textwidth]{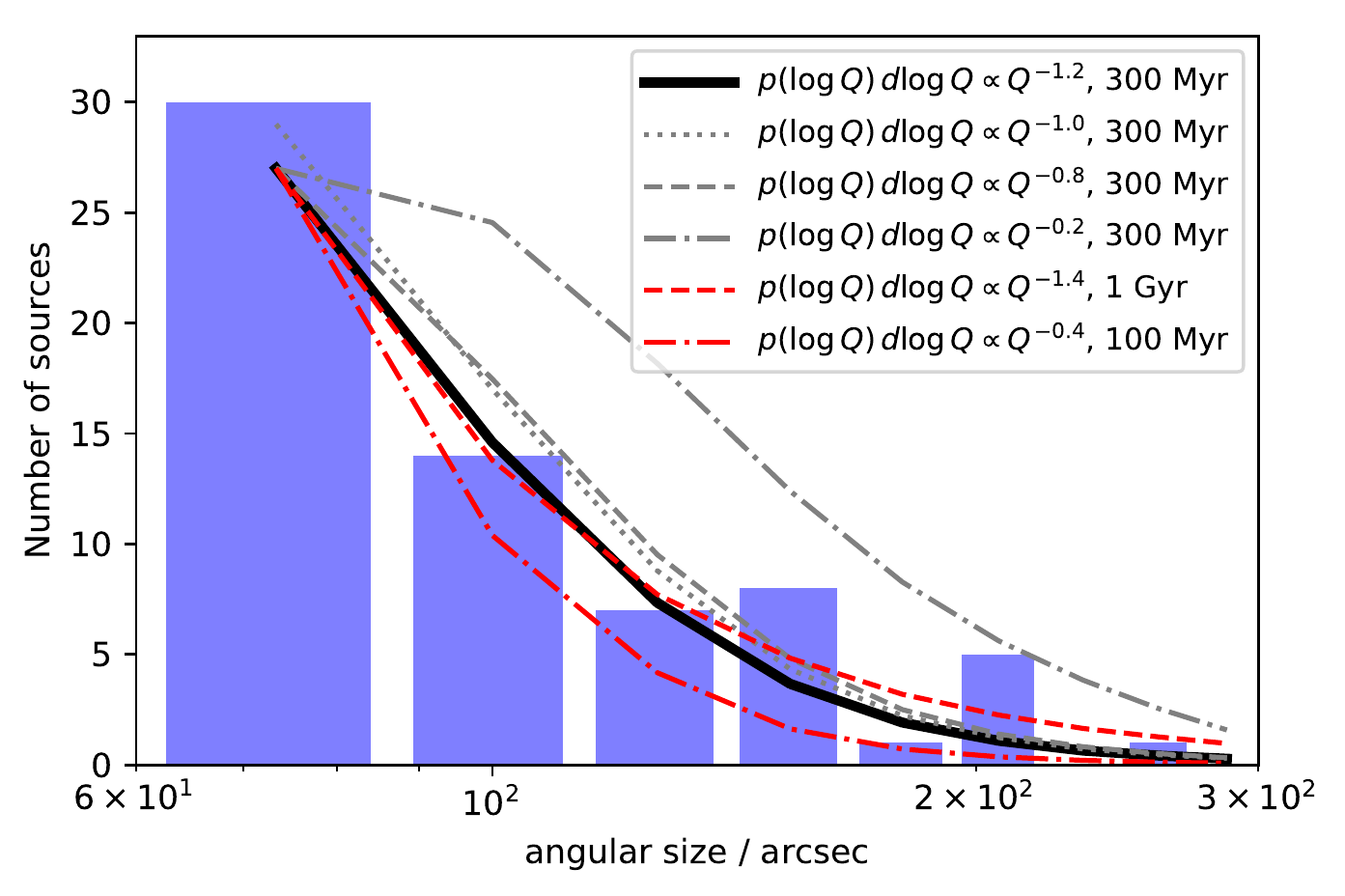}
\includegraphics[width=0.4\textwidth]{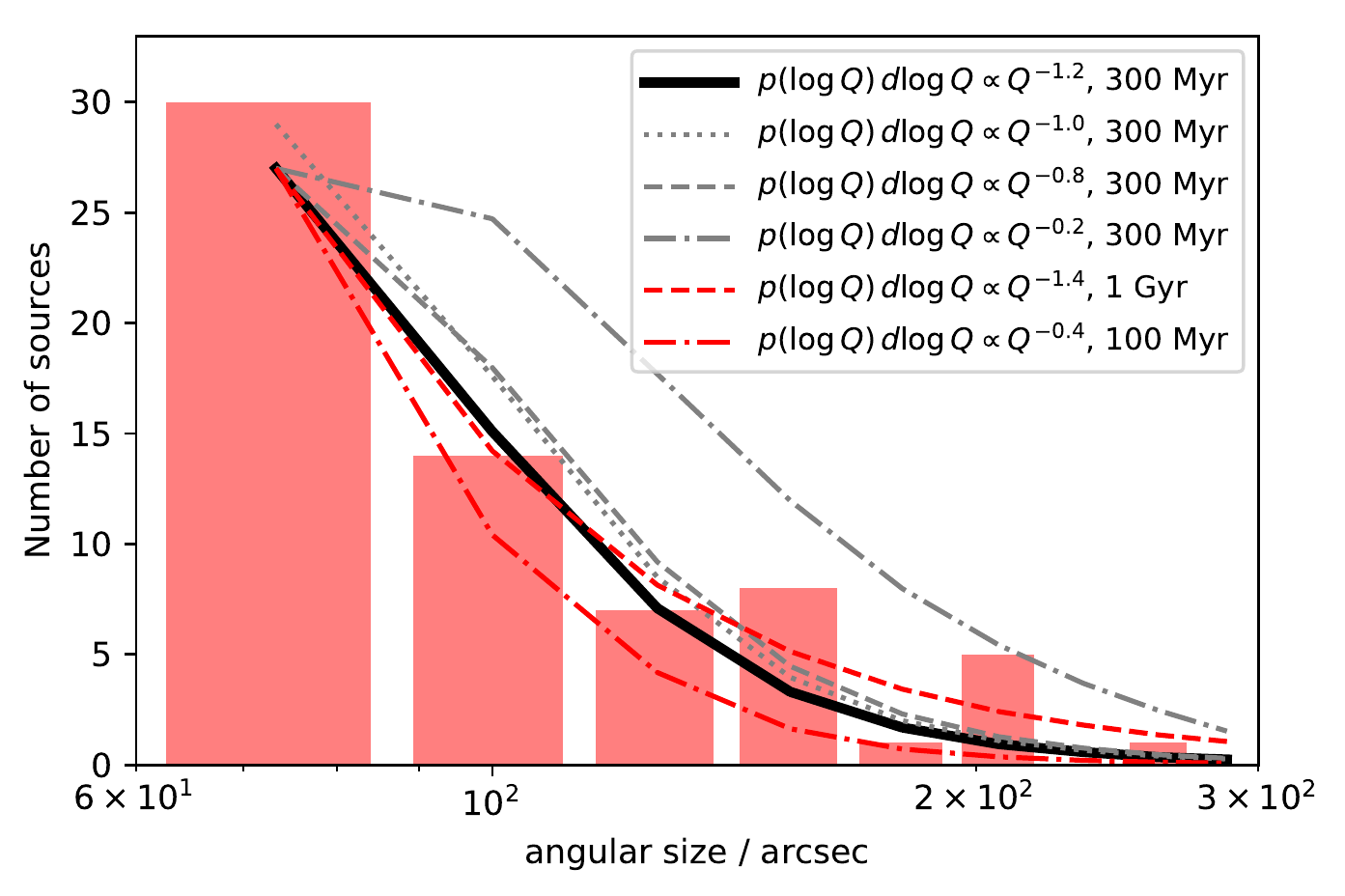}
\includegraphics[width=0.4\textwidth]{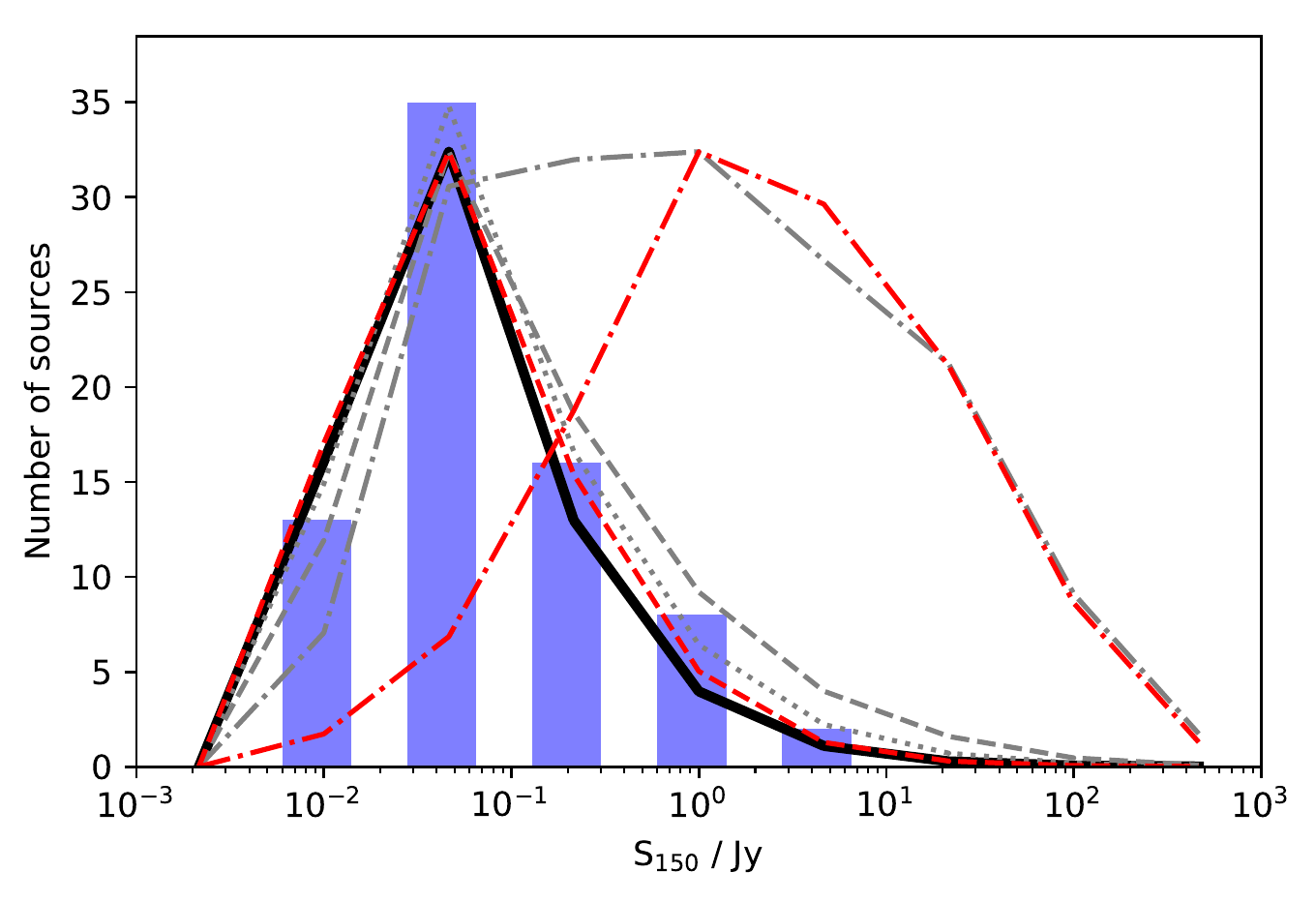}
\includegraphics[width=0.4\textwidth]{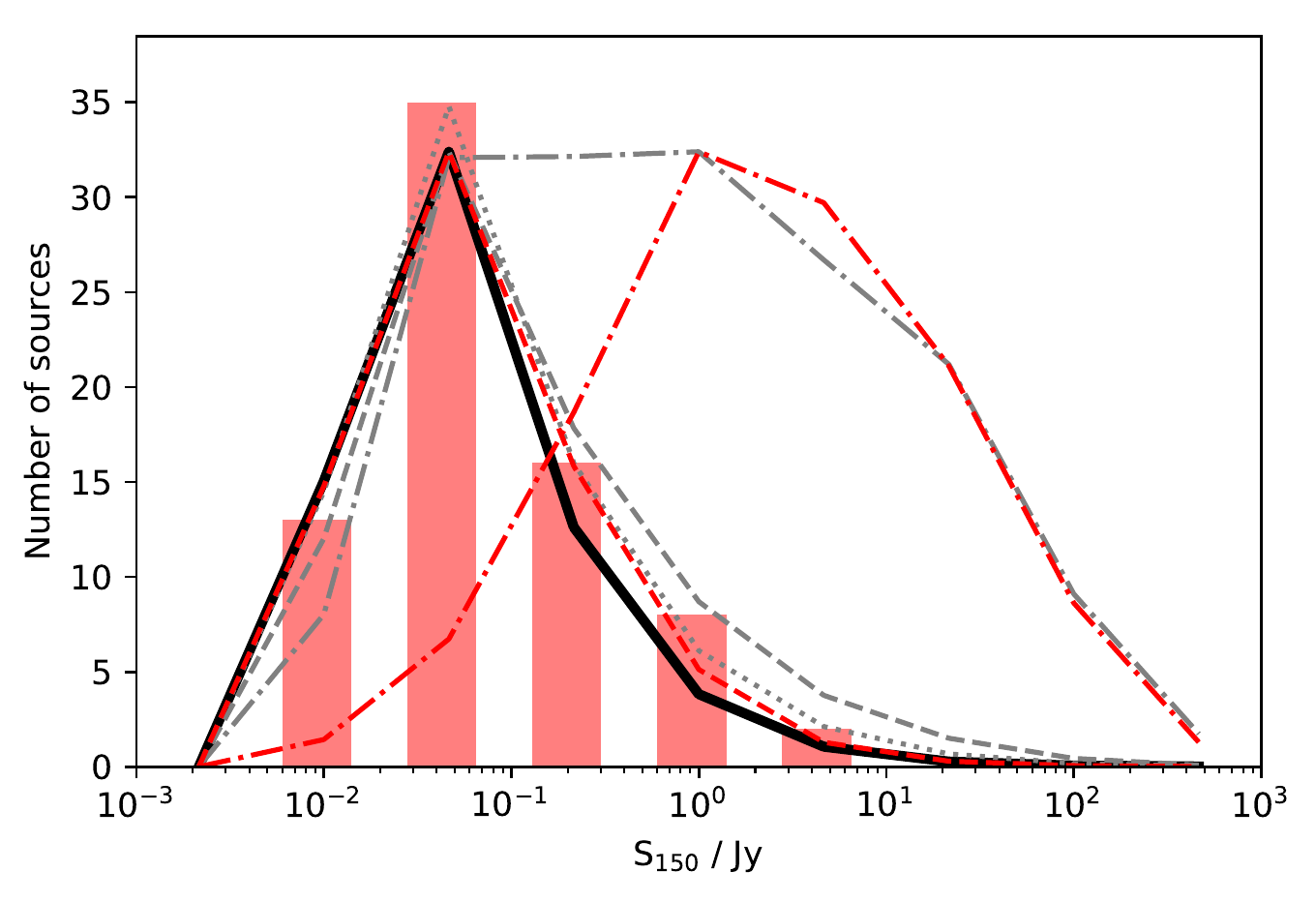}
\includegraphics[width=0.4\textwidth]{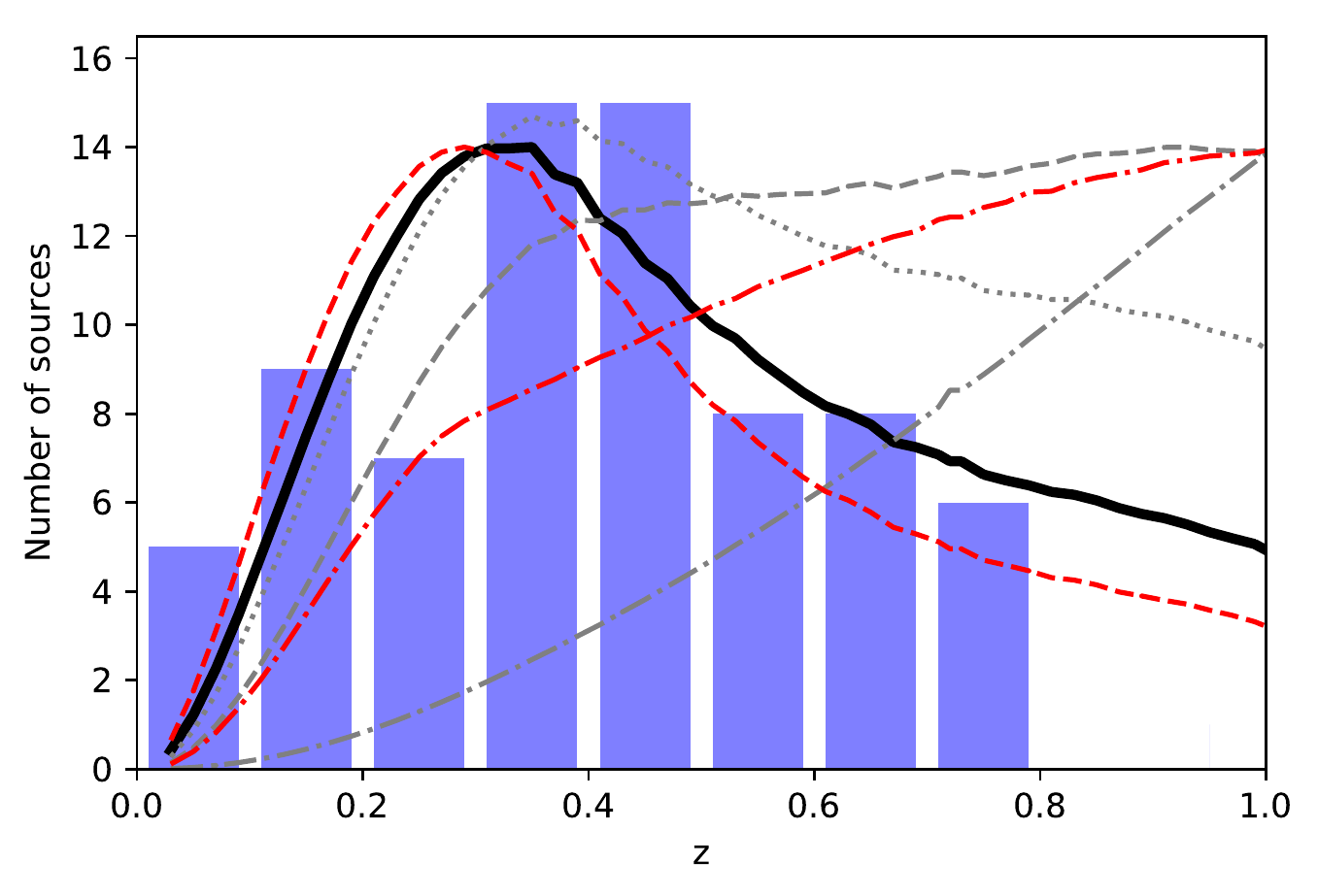}
\includegraphics[width=0.4\textwidth]{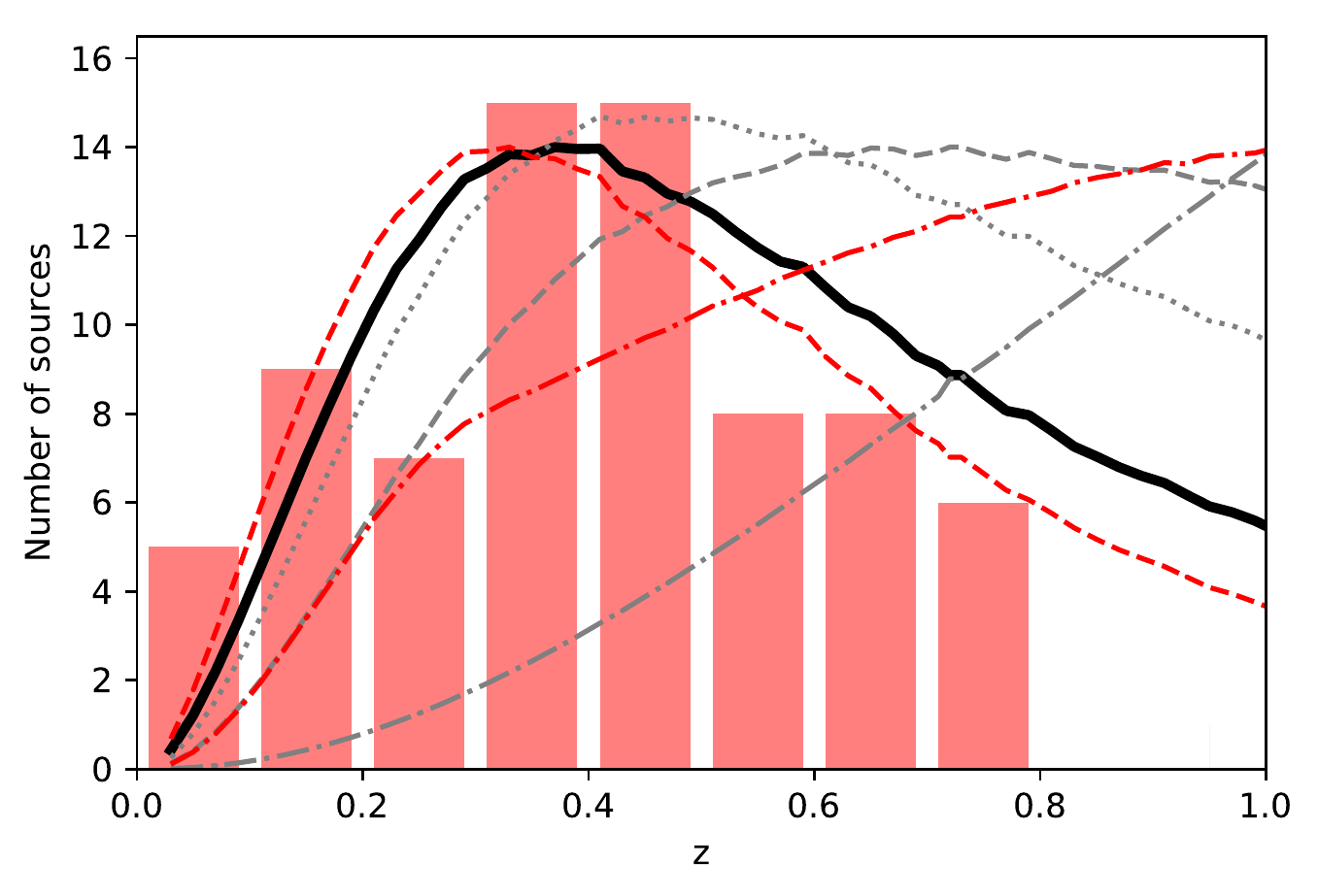}
\caption{Predicted distributions in angular size (top row), 150 MHz flux density (second row) and redshift (third row), for constant age models. Model predictions are shown as lines, with observations as histograms. The model normalisation is arbitrary. Left panels are for FR-I models, right panels for FR-II models. For both sets of models, comparison is made with the full sample of 74 active sources with redshifts. To fit the observed sizes and luminosities, models need to have power-law jet power indices $a=1.0-1.2$; short (100 Myr) and long (1 Gyr) sources cannot explain the observed distributions, consistent with the results of Figure~\ref{fig:samples_physical}. Best-fitting models are shown by black lines; for both FR-Is and FR-IIs, these correspond to $t_{\rm on}=300$~Myr and $p(\log Q) \, d \log Q \propto Q^{-1.2}$.}
\label{fig:pred_active_dist_best}
\end{center}
\end{figure*}

Figure~\ref{fig:pred_active_dist_best} shows the predicted observable population properties for a range of maximum source ages $t_{\rm on}$ and jet power distribution slopes $a$. A grid of RAiSE models in $(Q, z)$ is run to a maximum source age of $t_{\rm on}$, and output recorded every 1~Myr. At each timestep, we evaluate whether the expected angular size and surface brightness of the source would satisfy our sample selection criteria ($>60$~arcsec and $>200 \mu$Jy/beam, respectively). The fraction of time during which the source is detectable is then multiplied by the prior on input jet parameters, to yield a final prediction for the contribution of this part of parameter space to the observable population. We overplot the observed distributions for our extended source sample. Because of the relatively low-resolution of LOFAR, we cannot definitively assign FR-I or FR-II morphology for many sources; hence we use the full observed sample for comparison with models, noting that the FR-I and FR-II model tracks make very similar predictions.

For both FR-I and FR-II models, the best-fitting single age model has $a \geq 1.0$ and $t_{\rm on} \sim 300$~Myr. Younger and older ages cannot reproduce source sizes and flux densities simultaneously - for example, while the observed size distribution can be reproduced with shorter lifetimes and more powerful jets ($t_{\rm on}=100$ Myr, $a=0.4$), the flux density of the lobes is overestimated due to too many high-power sources; the opposite problem (too many large sources) occurs if the lifetime is too long. The fraction of compact sources observed by LOFAR provides an important constraint. \citet{HardcastleEA19} show that approximately two thirds of the most powerful radio sources (the progenitor population in this work) are resolved by LOFAR. In Table~\ref{tab:all_models} we present the expected fraction of sources more compact than $60$ arcseconds; models with many long-lived, low-power sources (e.g. $t_{\rm on}=1$ Gyr, $p(\log Q) d \log Q \propto Q^{-1.4}$) are ruled out by the observed compact fraction. In principle, the redshift distribution of observed sources should be a powerful discriminant between models; however in practice the association of radio sources to their host galaxies becomes increasingly more challenging at high redshift. We therefore cannot distinguish between models with similar slopes in the jet power distribution, such as the $a=1.0$ and $a=1.2$ models.% NOTE: HardcastleEA19 used 6 arcsec images, hence they should have a lower compact fraction.
%Shallow power-law indices also overpredict the number of high redshift sources. {\color{blue} We note that difficulties with identifying reliable optical counterparts mean the observed radio galaxy counts at high redshifts are almost certainly underestimates.} As we show in Section~\ref{sec:remnants}, constraints on $a$ place an important upper limit on the expect fraction of remnant sources.

\subsection{Power-law age models}
\label{sec:powerLawAgeModels}
% {\color{green} Taking sources older than 100 Myr (where we appear complete -- below this age there is a dearth of sources), observations are consistent with a power-law $N(t_{\rm on}) \propto t^{-1.3}$. Jet power is broadly consistent with $N(Q_{\rm jet}) \propto Q^{-1.0}$ or slightly flatter.}}

Our best single age models in Figure~\ref{fig:pred_active_dist_best} have $t_{\rm on} \sim 300$~Myr, broadly consistent with data fitting of individual objects, which are shown to be mostly younger than 1 Gyr. A more careful examination of Figure~\ref{fig:samples_physical}, however, shows that the observed age distribution of old ($>100$~Myr, where we are complete) sources is consistent with a declining power law, approximately $p(\log t_{\rm on}) d \log t_{\rm on} \propto t^{-1}$. This is qualitatively consistent with expectation from simulations of feedback-regulated black hole accretion \citep{NovakEA11,GaspariEA17}, which show that black hole accretion rates follow a power spectrum consistent with pink noise.

In our second set of models, we therefore adopt a power-law distribution in source age, in addition to a power-law distribution in jet power. In our models, we assume that these quantities are not correlated; this would be expected if the two distributions were largely set by different processes, for example black hole spin \citep{Daly09,Daly16} for jet power and black hole accretion \citep{NovakEA11,GaborBournaud13} for the duty cycle.

\begin{figure*}
\begin{center}
\includegraphics[width=0.4\textwidth]{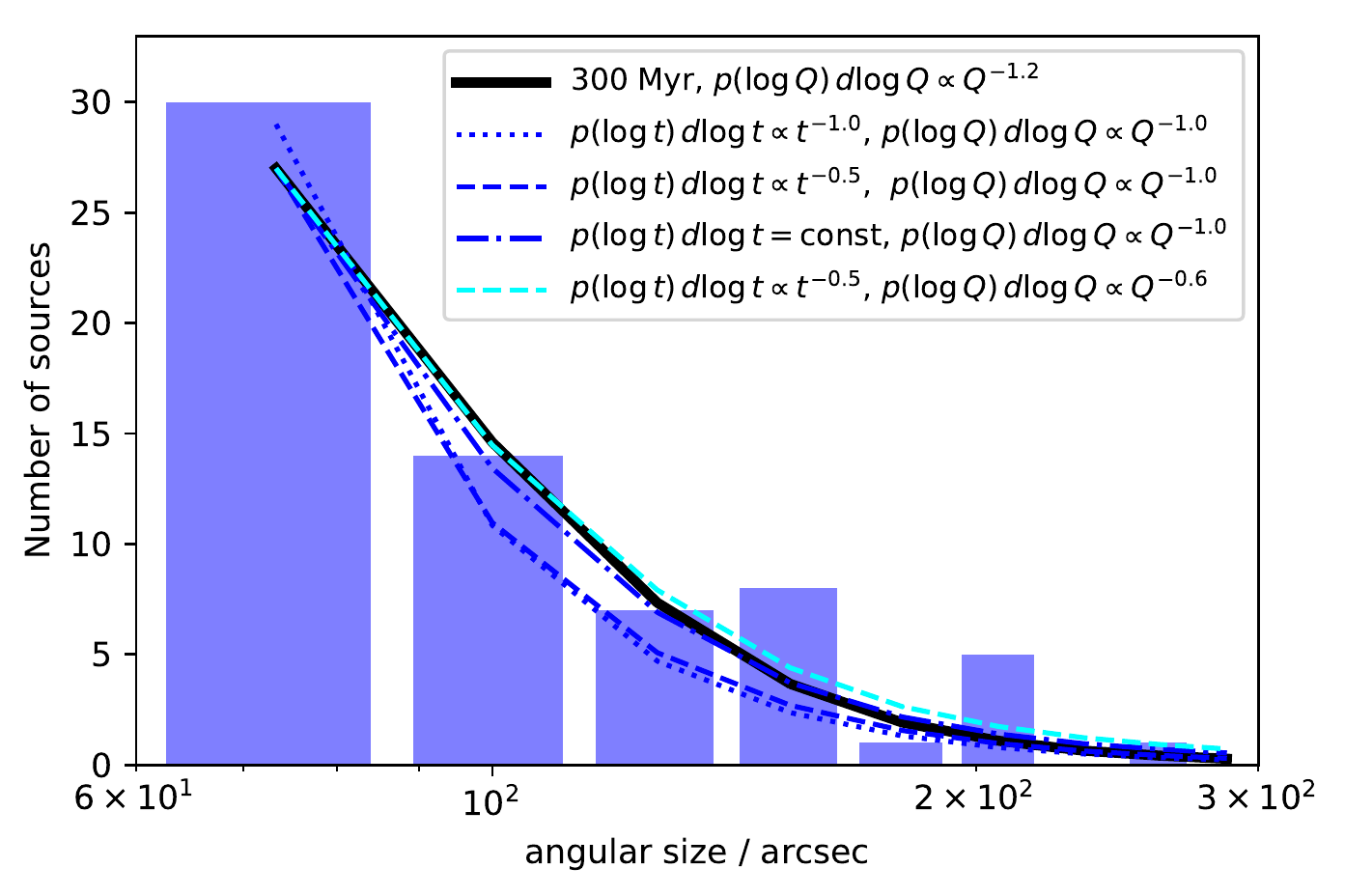}
\includegraphics[width=0.4\textwidth]{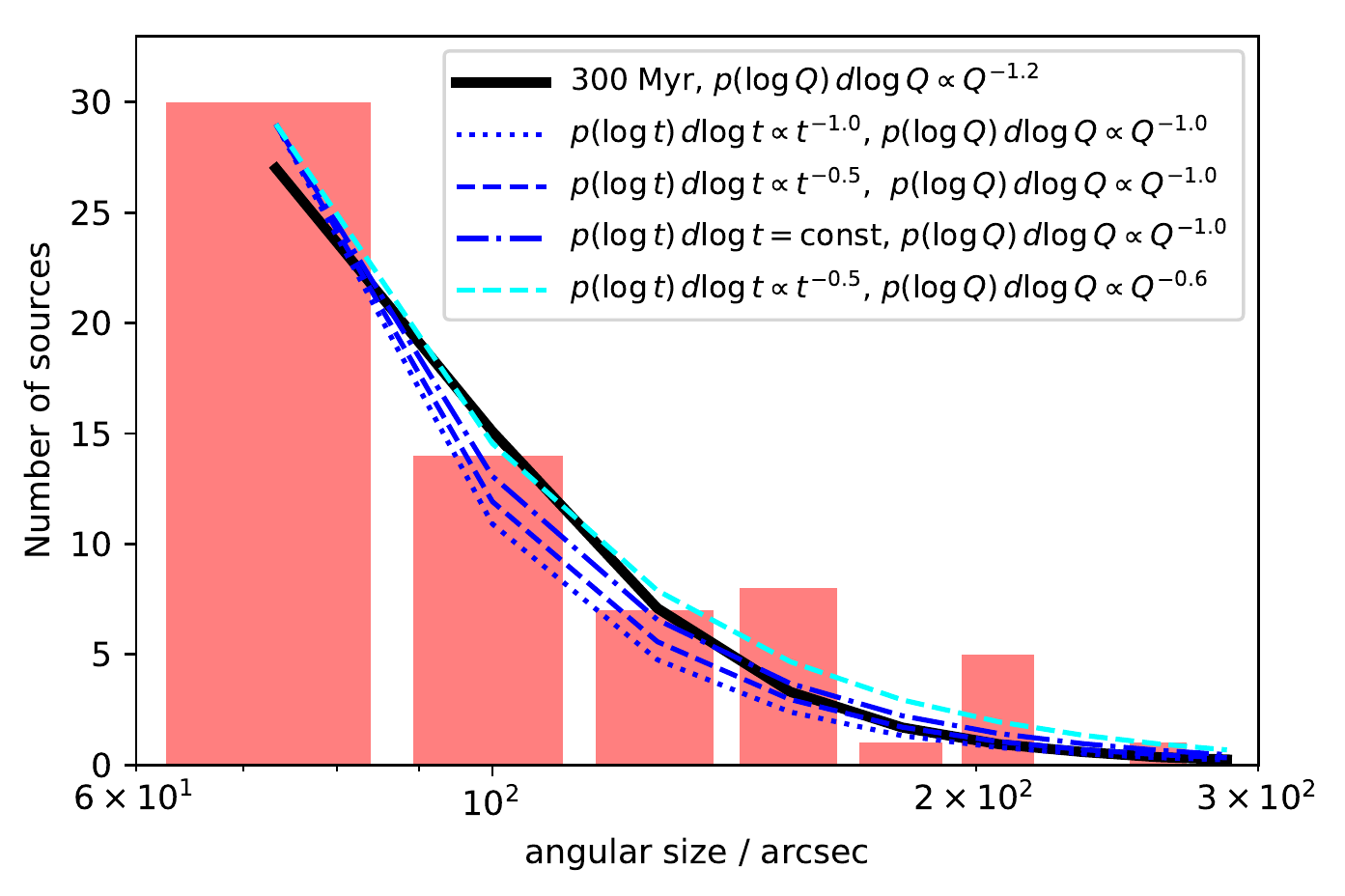}
\includegraphics[width=0.4\textwidth]{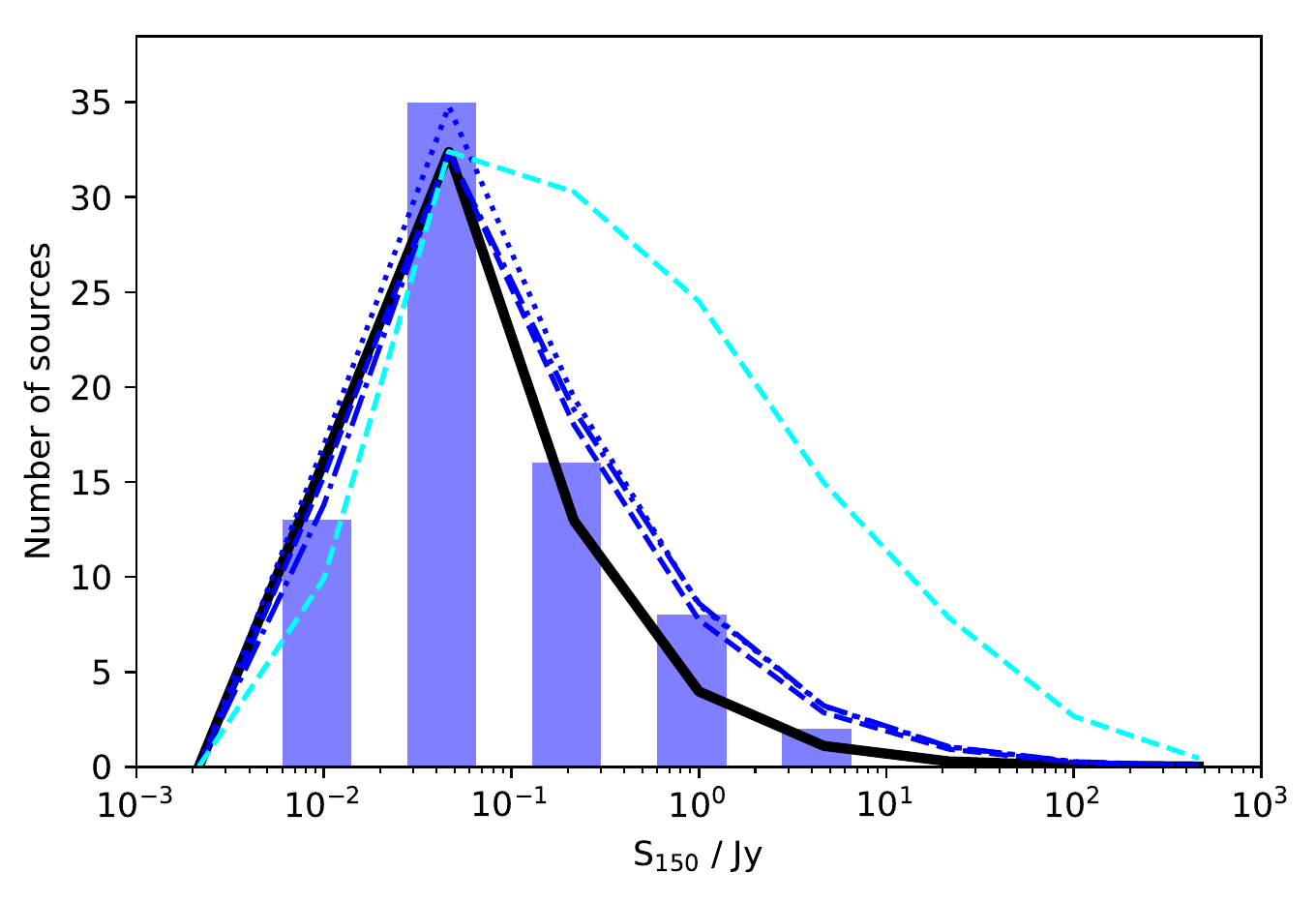}
\includegraphics[width=0.4\textwidth]{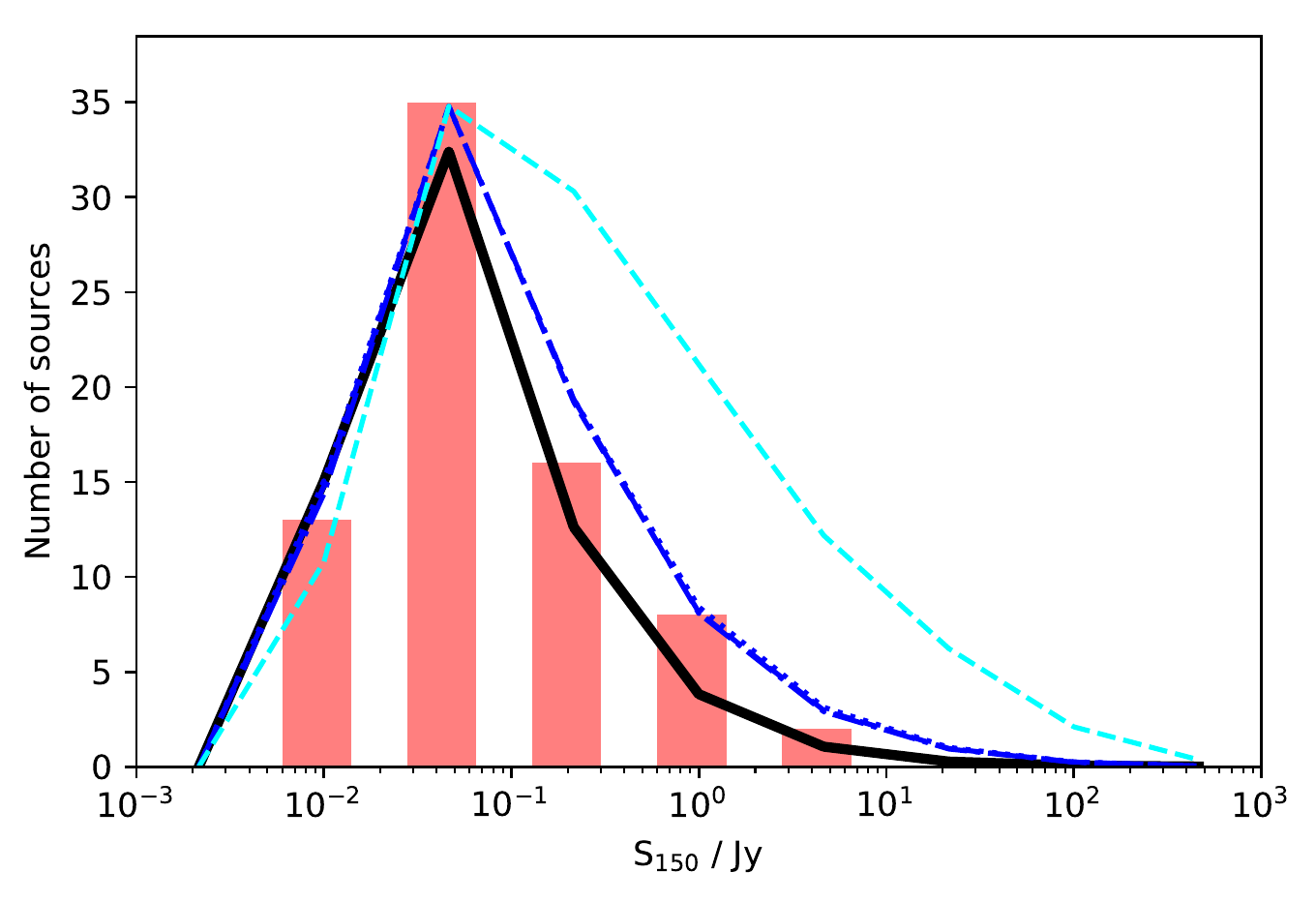}
\includegraphics[width=0.4\textwidth]{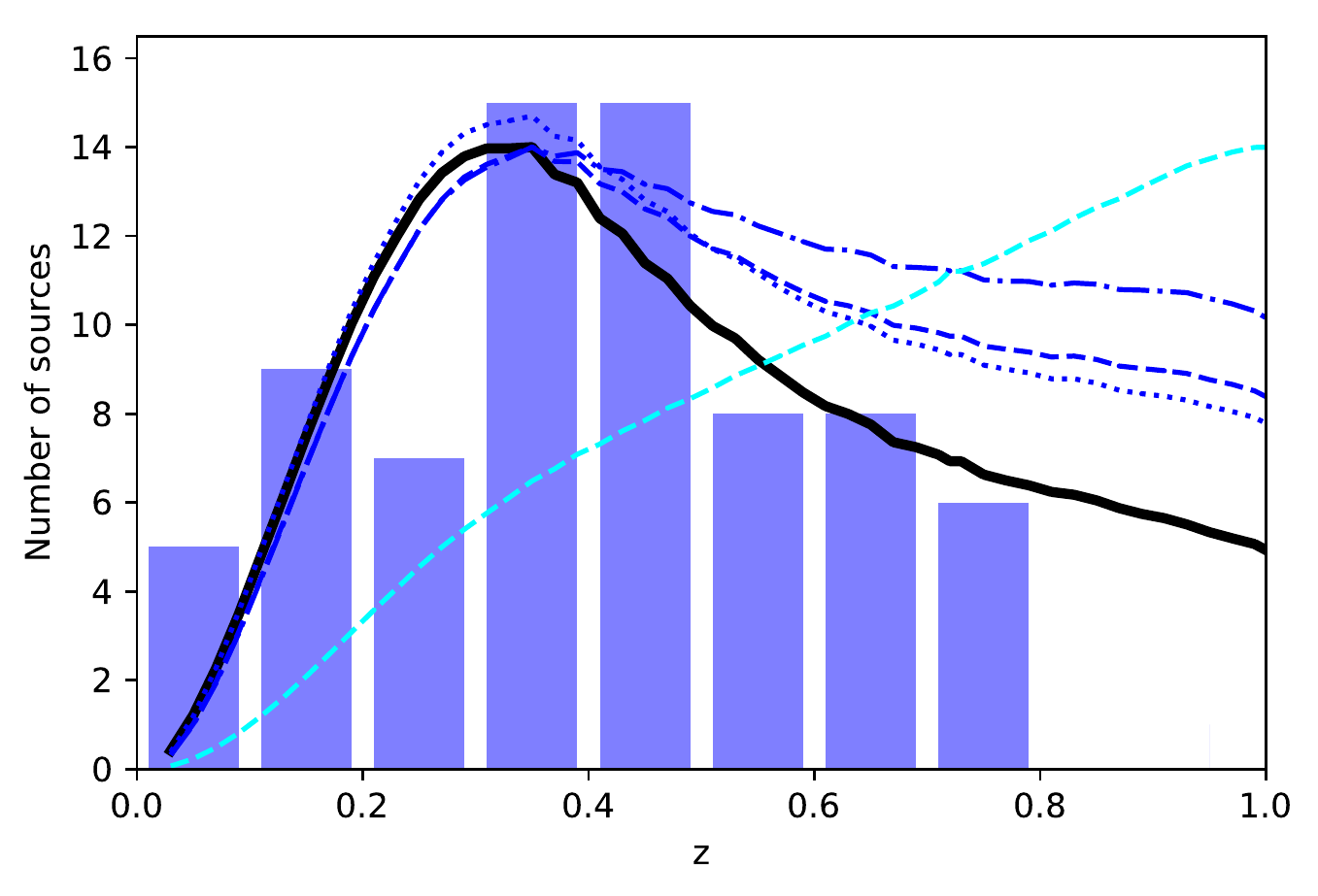}
\includegraphics[width=0.4\textwidth]{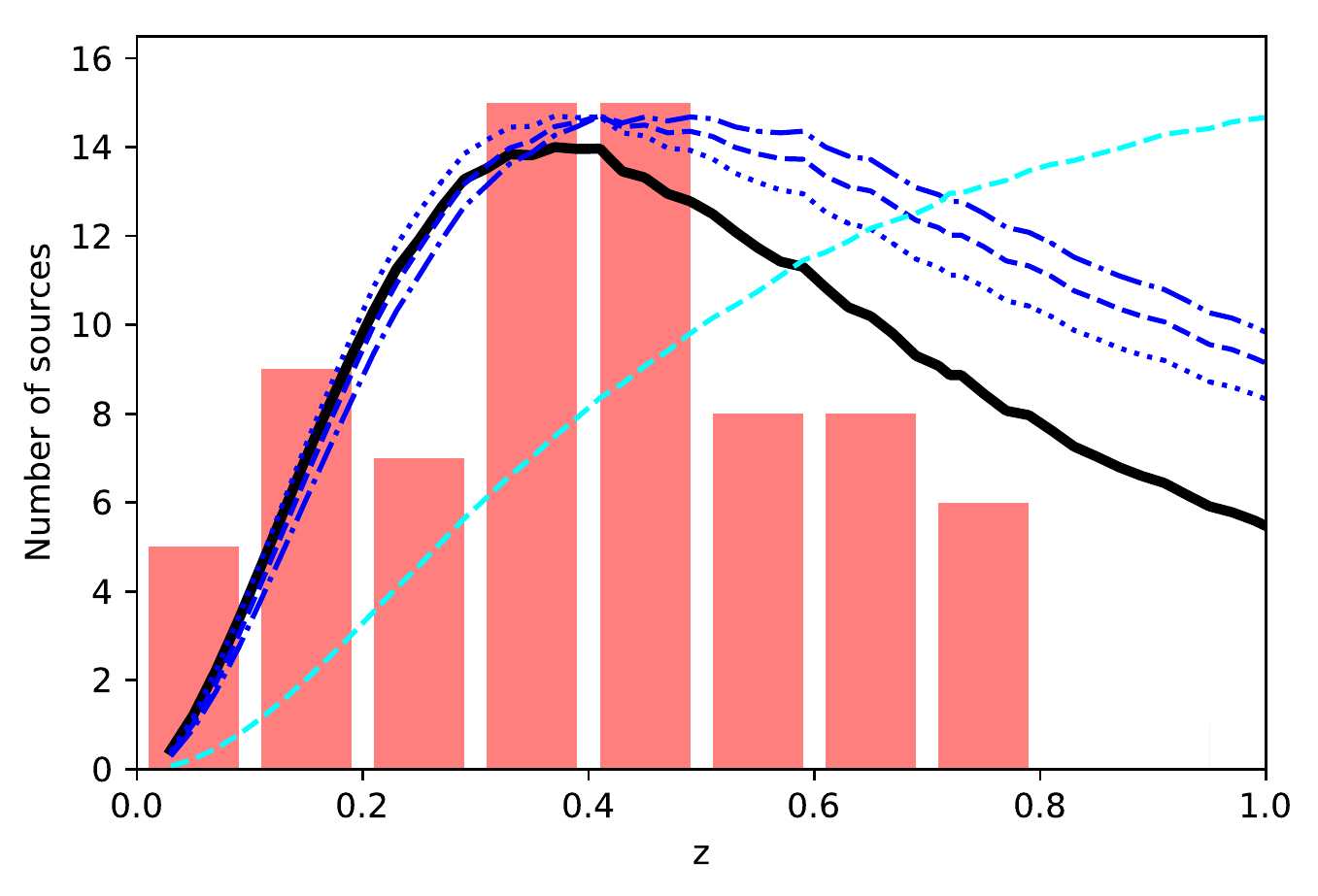}
\caption{Predicted distributions in angular size (top row), 150 MHz flux density (second row), and redshift (third row), for power-law age models. As in Figure~\ref{fig:pred_active_dist_best}, histograms show observational constraints, and lines represent model predictions. Left panels are for FR-I models, right panels for FR-II models. Best-fit single age model (Section~\ref{sec:constantAgeModels}) predictions are shown as black solid lines. A range of negative power-law slopes (between -0.5 and -1.5) provides suitable fits to data.}
\label{fig:pred_active_dist_best_plAge}
\end{center}
\end{figure*}

Figure~\ref{fig:pred_active_dist_best_plAge} shows that a range of plausible power-law age exponents is consistent with the observed properties of the active radio galaxy population. Hence, both single age and the (more complex) age distribution models can explain the observed properties of the active sources. However, as first pointed out by \citet{Hardcastle18} and shown in detail in the following section, these two sets of models make very different predictions for the remnant and restarted source populations.

%While a distribution of ages may be more realistic, we prefer to avoid introducing poorly constrained parameters (describing the shape of this distribution) into our models. Recently, dynamical modelling of 23,344 radio-loud AGN in the LOFAR field by \citet{HardcastleEA19} showed that models with a uniform age distribution gave better agreement with the data than log-uniform distributions, with the latter underpredicting the sizes of bright sources in the range considered in this work ($10^{25} \leq L_{150} \leq 10^{27}$~W/Hz). Our derived maximum source age is broadly consistent with the estimate of 1~Gyr by \citet{HardcastleEA19}; the quantitative difference is due to detailed model assumptions, most likely on jet environments. Comparison of the left and right panels of Figure~\ref{fig:pred_active_dist_best} shows that there is very little difference in the predicted population properties for FR-II and FR-I models. Furthermore, by construction our remnant models (below) are adjusted to reproduce the observed properties of active radio galaxies, their presumed progenitor population. Hence, we do not expect uncertainties in the modelling (i.e. FR-I or FR-II jets, environments) to dominate our remnant predictions in the following section.

 %{\color{red} This analysis assumes that there are no hidden systematic effects against selecting high-redshift sources.} 

\begin{table*}															
\centering															
%\tiny		
\begin{adjustbox}{width=\linewidth,center}													
\tabcolsep=0.4cm															
\begin{tabular}{ c | c  | c c | c }															
\hline\hline															
						
& \multicolumn{1}{|c|}{Model}															
& \multicolumn{2}{|c|}{Compact fraction}															
 & \multicolumn{1}{|c}{Comment}\\															
& &	FR-I	&	FR-II	&	\\
\hline\hline															
& $t_{\rm on}=300$ Myr, $p(\log Q) d \log Q \propto Q^{-1.2}$	&	0.74	&	0.75	&	best model	\\
Single age & $t_{\rm on}=300$ Myr, $p(\log Q) d \log Q \propto Q^{-1.0}$	&	0.78	&	0.79	&	good fit	\\
& $t_{\rm on}=300$ Myr, $p(\log Q) d \log Q \propto Q^{-0.8}$	&	0.68	&	0.68	&	good fit; compact fraction may be too low \\
\hline															
& $t_{\rm on}=300$ Myr, $p(\log Q) d \log Q \propto Q^{-0.2}$	&	0.37	&	0.37	&	too many bright and large sources\\
& $t_{\rm on}=1$ Gyr, $p(\log Q) d \log Q \propto Q^{-1.4}$	&	0.51	&	0.52	&	compact fraction too low	\\
& $t_{\rm on}=100$ Myr, $p(\log Q) d \log Q \propto Q^{-0.4}$	&	0.79	&	0.80	&	too many bright sources	\\
\hline\hline															
Power-law age & $p(t_{\rm on}) \propto t_{\rm on}^{-1.0}$, $p(\log Q) d \log Q \propto Q^{-1.0}$	&	0.74	&	0.82	&	best model	\\
& $p(\log t_{\rm on}) d \log t_{\rm on} \propto t_{\rm on}^{-0.5}$, $p(\log Q) d \log Q \propto Q^{-1.0}$	&	0.71	&	0.72	&	good fit	\\
\hline															
& $p(\log t_{\rm on}) d \log t_{\rm on} \propto t_{\rm on}^{0}$, $p(\log Q) d \log Q \propto Q^{-1.0}$	&	0.61	&	0.62	&	compact fraction too low\\
& $p(\log t_{\rm on}) d \log t_{\rm on} \propto t_{\rm on}^{-0.5}$, $p(\log Q) d \log Q \propto Q^{-0.6}$	&	0.56	&	0.56	&	too many bright and high-z sources; compact fraction too low\\

\hline															
\end{tabular}	
\end{adjustbox}														
\caption{Compact fractions predicted by dynamical models. All plausible single-age ($t_{\rm on}=300$~Myr, jet power exponent between $-0.8$ and $-1.2$) and power-law age (exponent between $-0.5$ and $-1.0$) models predict compact fractions ($<60$ arcsec) of between 0.68-0.82. FR-I and FR-II model predictions are consistent.}
\label{tab:all_models}															
\end{table*}																																			 
 
\section{Predicted remnant and restarted fractions}
\label{sec:remnants}

We use the above models to make predictions for the remnant and restarted populations. In our models, we evolve radio sources in their active phase until they switch off; after this point we evolve the lobes as remnants (e.g. Figure~\ref{fig:example_tracks}) until they fade below the detection limit. During the remnant phase, the black hole activity may re-start again; if the lobes are still visible above the LOFAR surface brightness detection limit, we expect to detect a restarted source. In our models, radio emission from the second (young) radio burst is not explicitly modelled; this assumption is justified by \citet[][]{JurlinEA20}'s finding that these second bursts are overwhelmingly compact, and the overall luminosity at LOFAR frequencies is dominated by the diffuse lobe emission. In our analysis below, we combine the remnant and restarted source populations when comparing with model predictions.

\subsection{Constant age models}

\begin{figure*}
\begin{center}
\includegraphics[width=0.4\textwidth]{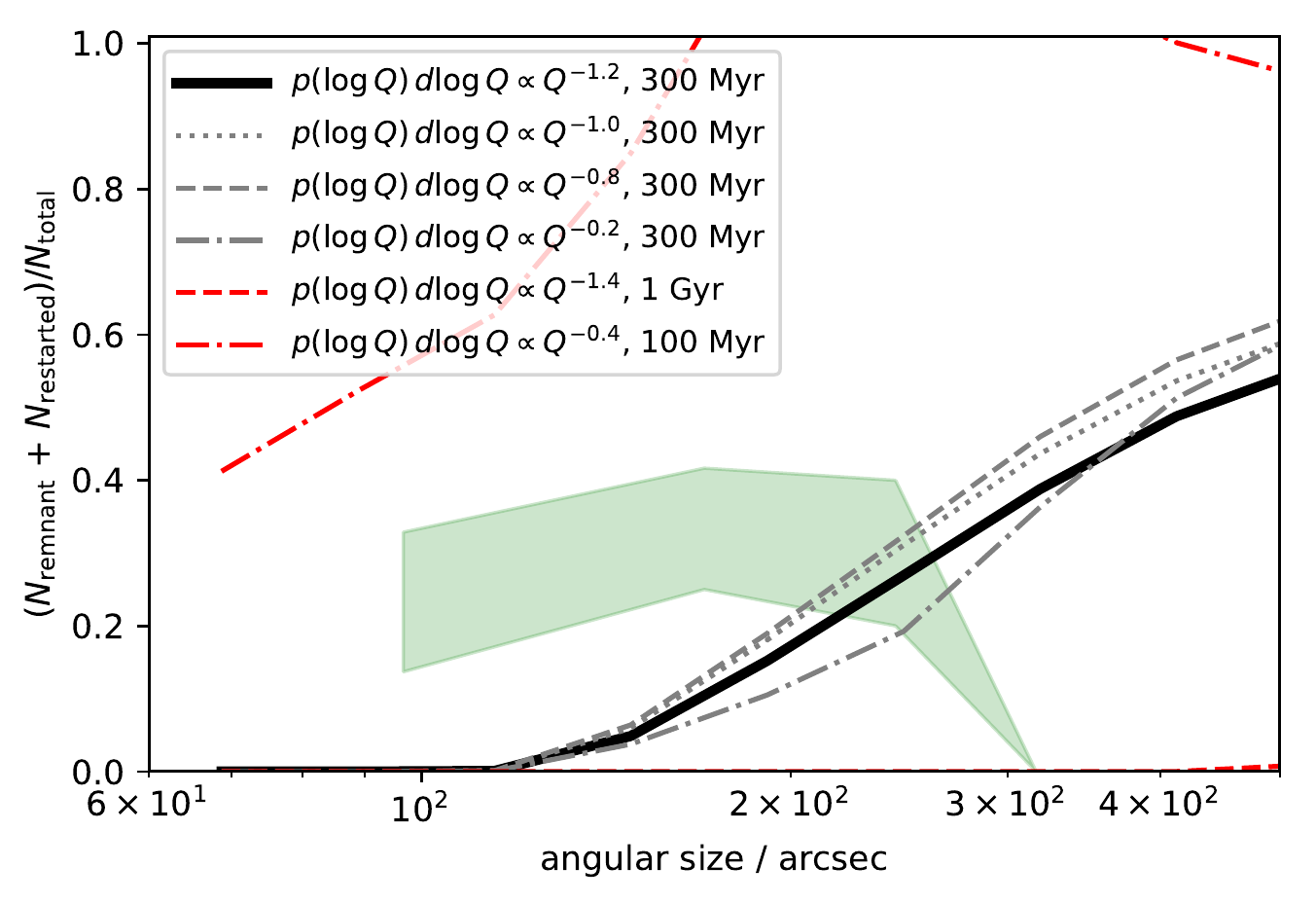}
\includegraphics[width=0.4\textwidth]{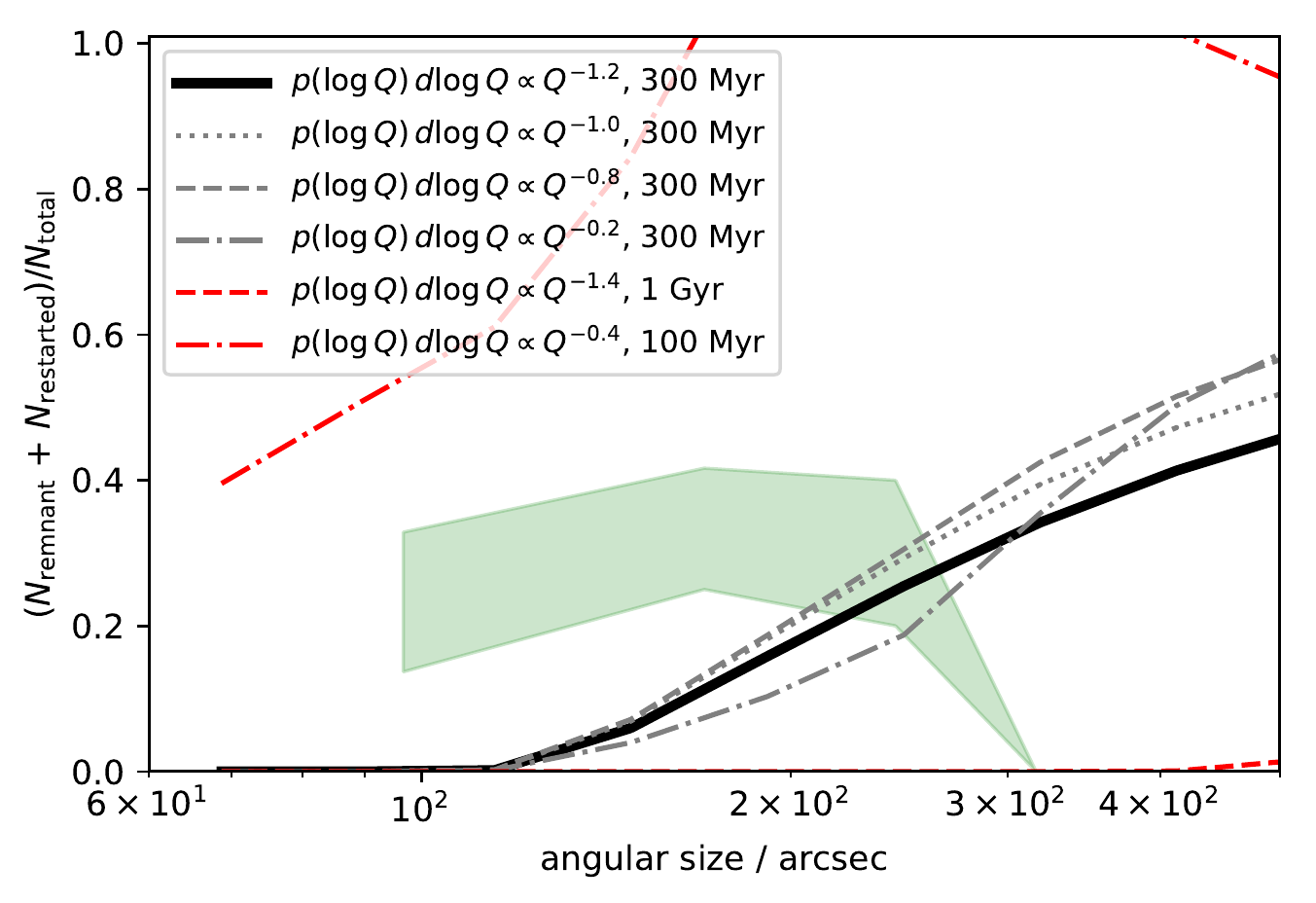}
\includegraphics[width=0.4\textwidth]{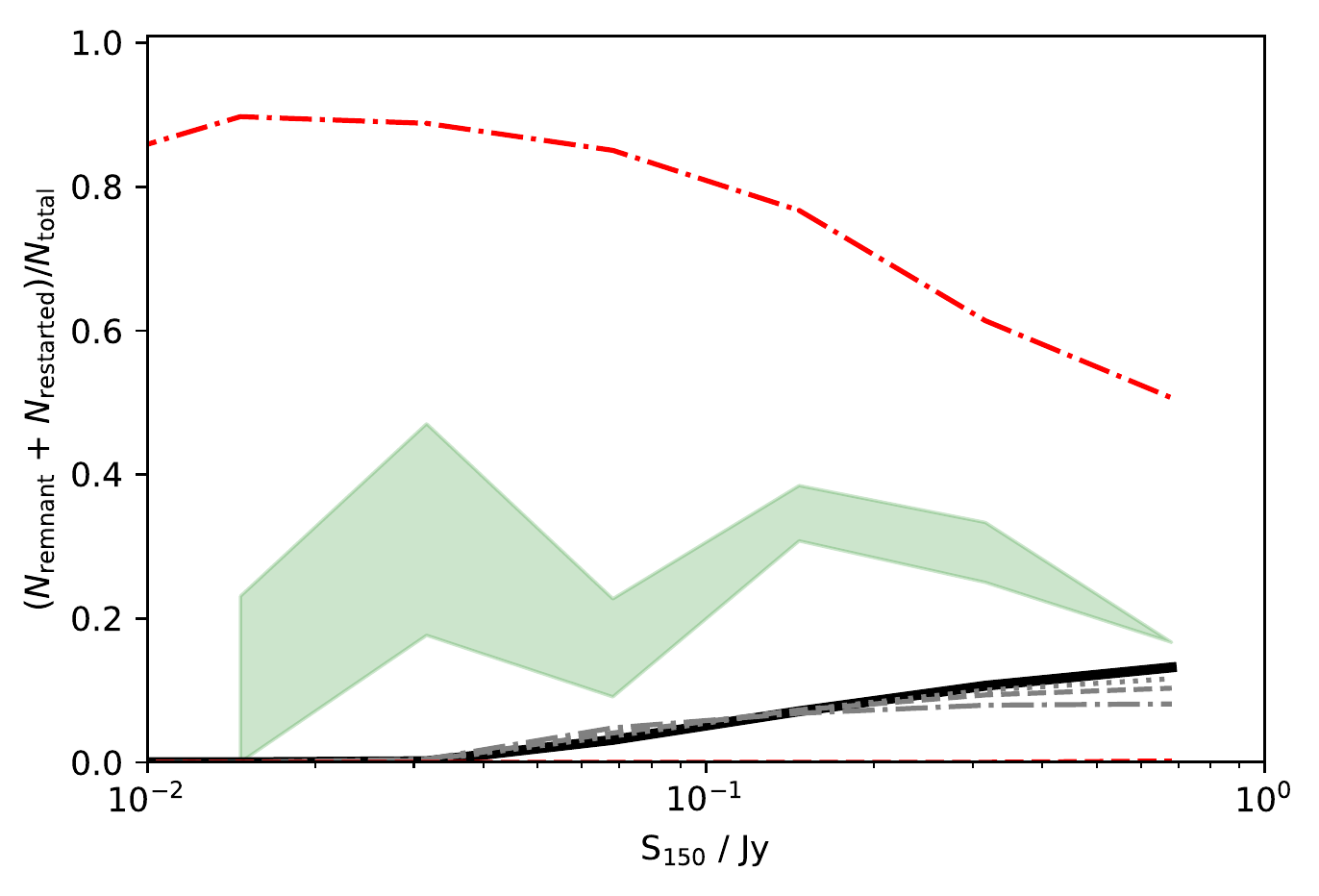}
\includegraphics[width=0.4\textwidth]{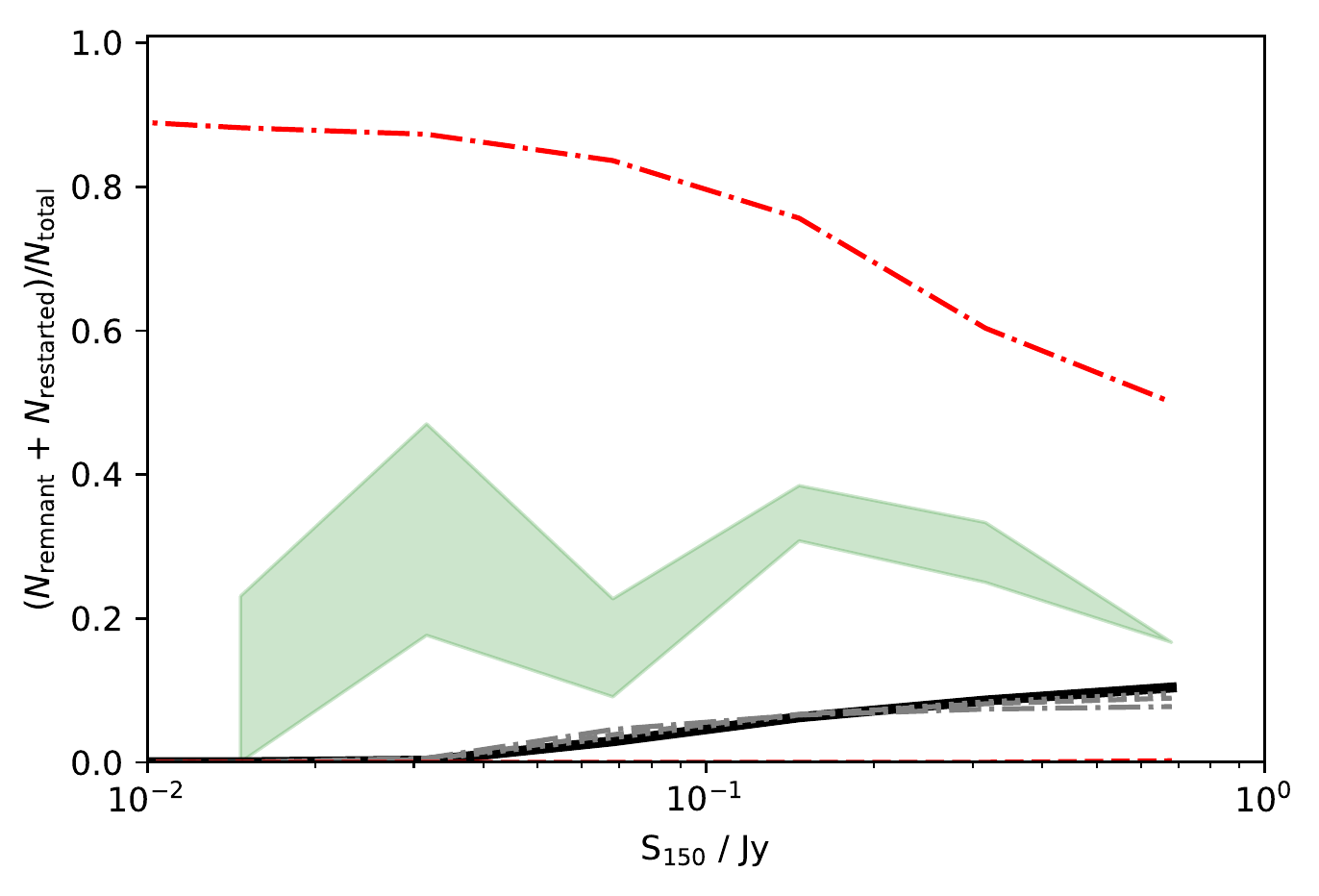}
\includegraphics[width=0.4\textwidth]{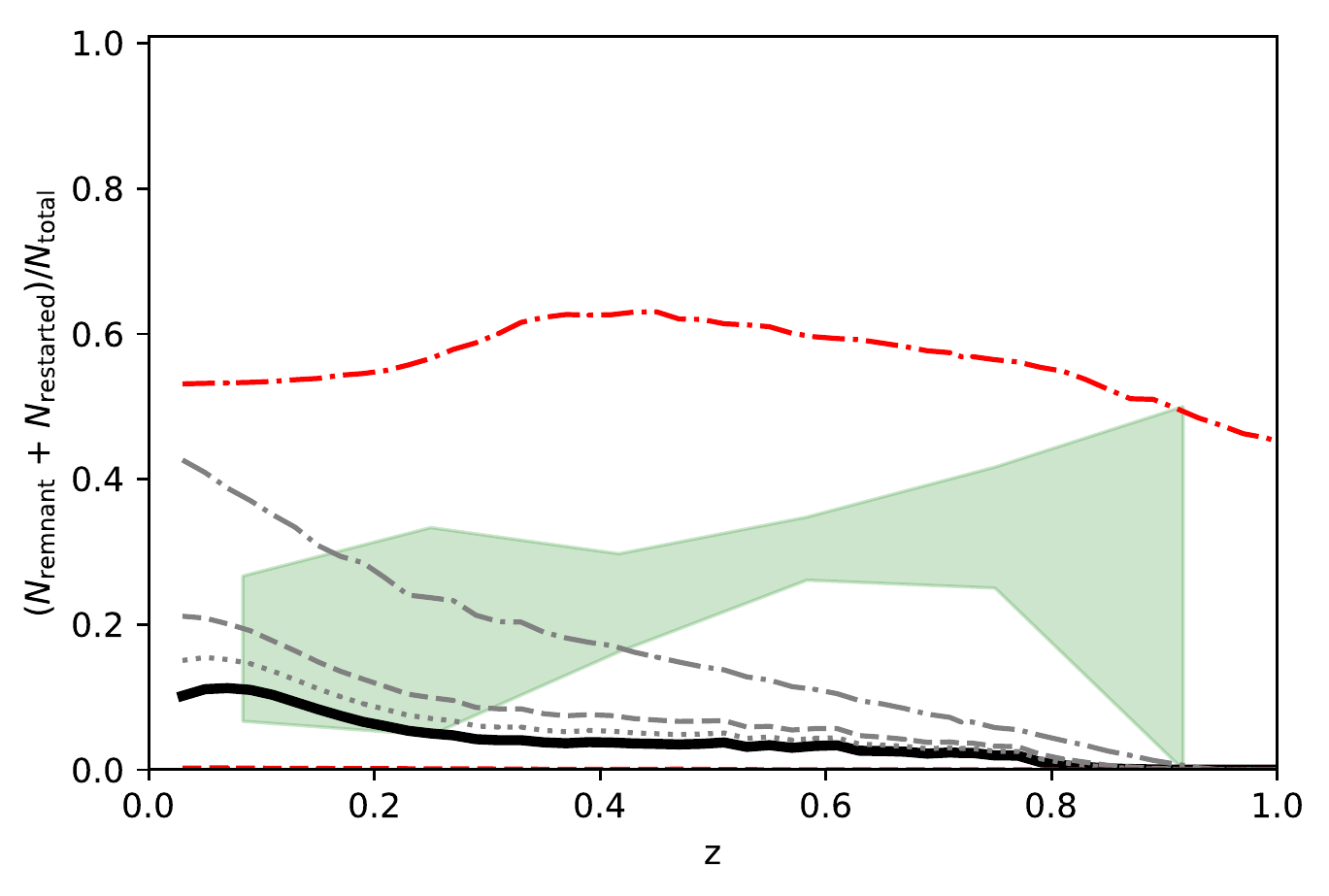}
\includegraphics[width=0.4\textwidth]{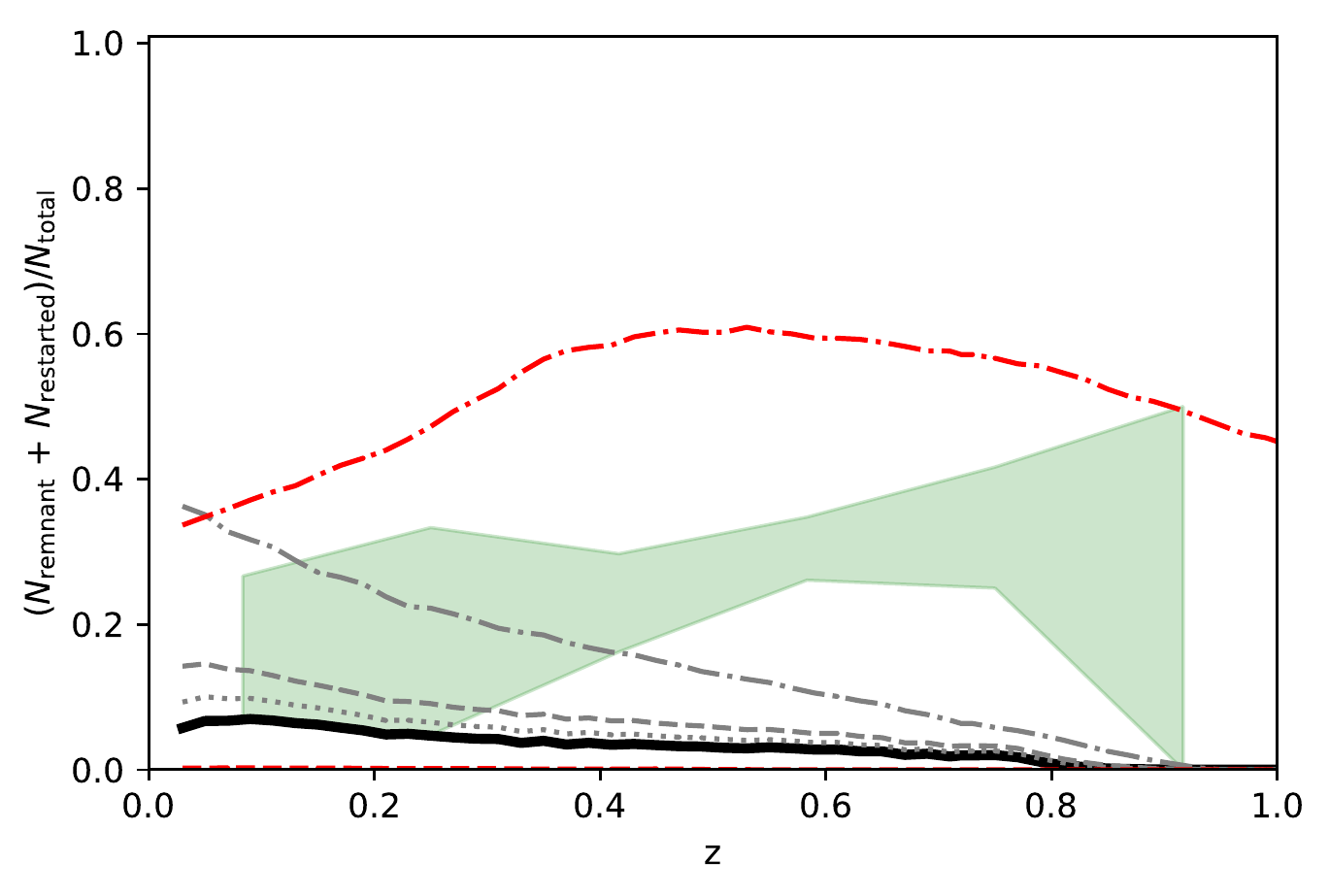}
\caption{Predicted remnant fractions for FR-Is (left) and FR-IIs (right) as a function of observable parameters, for constant age models. Remnant plus restarted fraction is predicted to decline with redshift, and increase with flux density and source size. Green shaded regions are observational constraints from the sample of \citet{JurlinEA20}.%Almost all very steep spectrum sources should be remnants, but very few of these are expected to be found as remnants fade rapidly once the jets are switched off.
}
\label{fig:pred_remnantFrac_constAge}
\end{center}
\end{figure*}

Figure~\ref{fig:pred_remnantFrac_constAge} shows the predicted remnant and restarted fractions as a function of observable parameters, for constant age models. Shaded regions in Figure~\ref{fig:pred_remnantFrac_constAge} show observational constraints from the LOFAR Lockman Hole sample of candidate remnant and restarted sources. As discussed in \citet{JurlinEA20}, confirming candidate restarted sources is challenging: these are selected based on a combination of core prominence, steep core spectral index, and/or visual morphology characteristic of double-double radio sources; still, some candidate restarted sources may in fact be ``normal'' active radio galaxies with bright cores. We therefore calculate two constraints from observations: the upper limit on the remnant and restarted fraction is obtained by assuming all candidate restarted sources are classified correctly; and the lower limit by assuming none of them is (i.e. all candidate restarted sources are in fact normal radio galaxies) except for the two double-double radio galaxies. This approach implicitly assumes that our remnant classification is robust.

Once integrated over the observables (flux densities, sizes and redshifts), all plausible single age models ($t_{\rm on}=300$~Myr, $a=0.8-1.2$) predict remnant plus restarted fractions of only between 2 and 5 percent. Observationally, the lower limit on the remnant plus restarted fraction (obtained from remnants alone) is $\geq 9$ percent percent \citet{MahatmaEA18} or $\geq 11$ \citet{JurlinEA20}. This alone does not rule out single-age models. However, single-age models fail to explain the observed statistics of restarted sources: the fraction of double-double radio sources alone is $\geq 4$ percent \citep{MahatmaEA19}; and the total restarted fraction is likely much higher than this \citep[e.g. 13-15 percent reported by ][]{JurlinEA20}. Hence, single-age models appear in tension with these data.

\subsection{Power-law age models}

\begin{figure*}
\begin{center}
\includegraphics[width=0.4\textwidth]{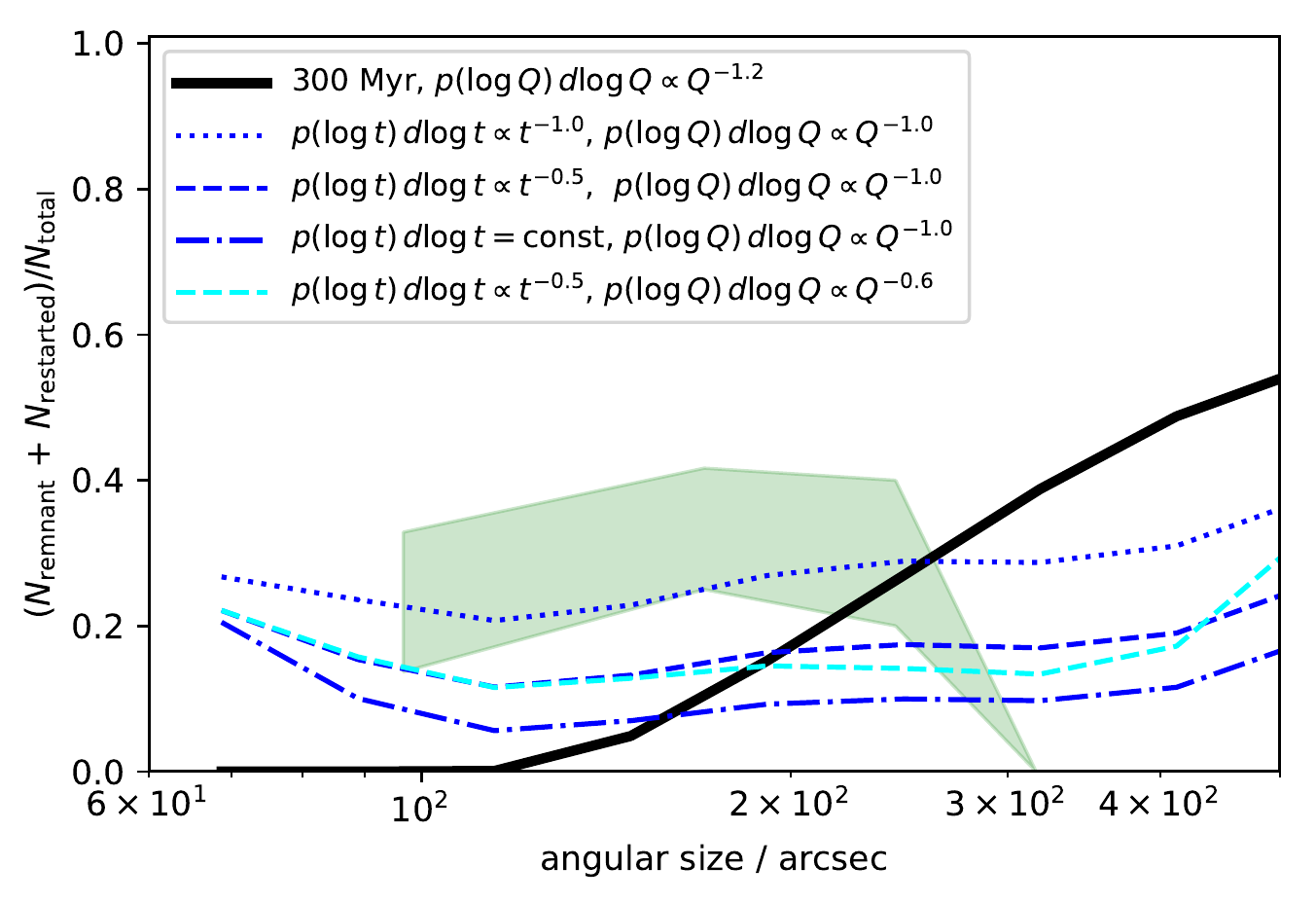}
\includegraphics[width=0.4\textwidth]{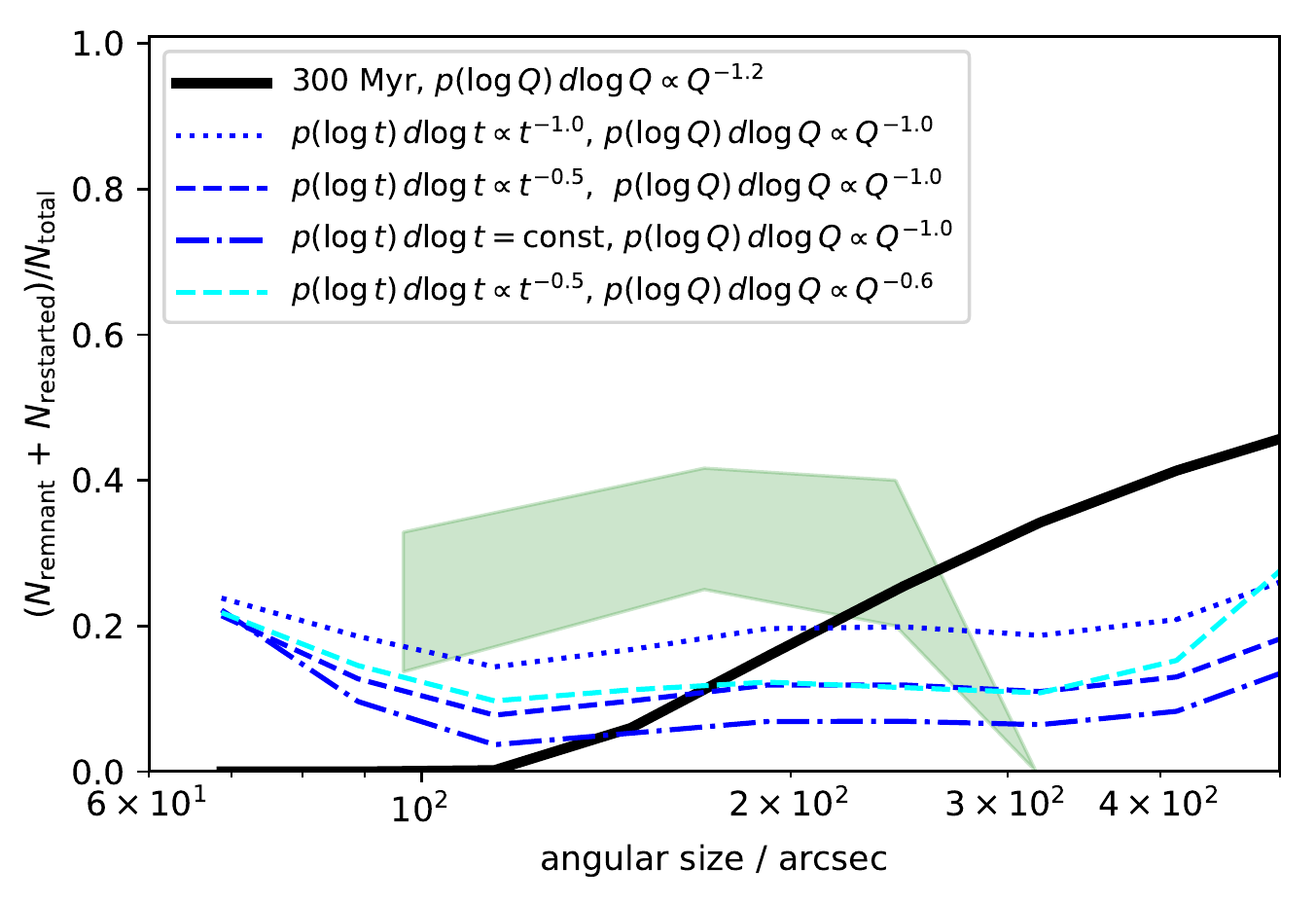}
\includegraphics[width=0.4\textwidth]{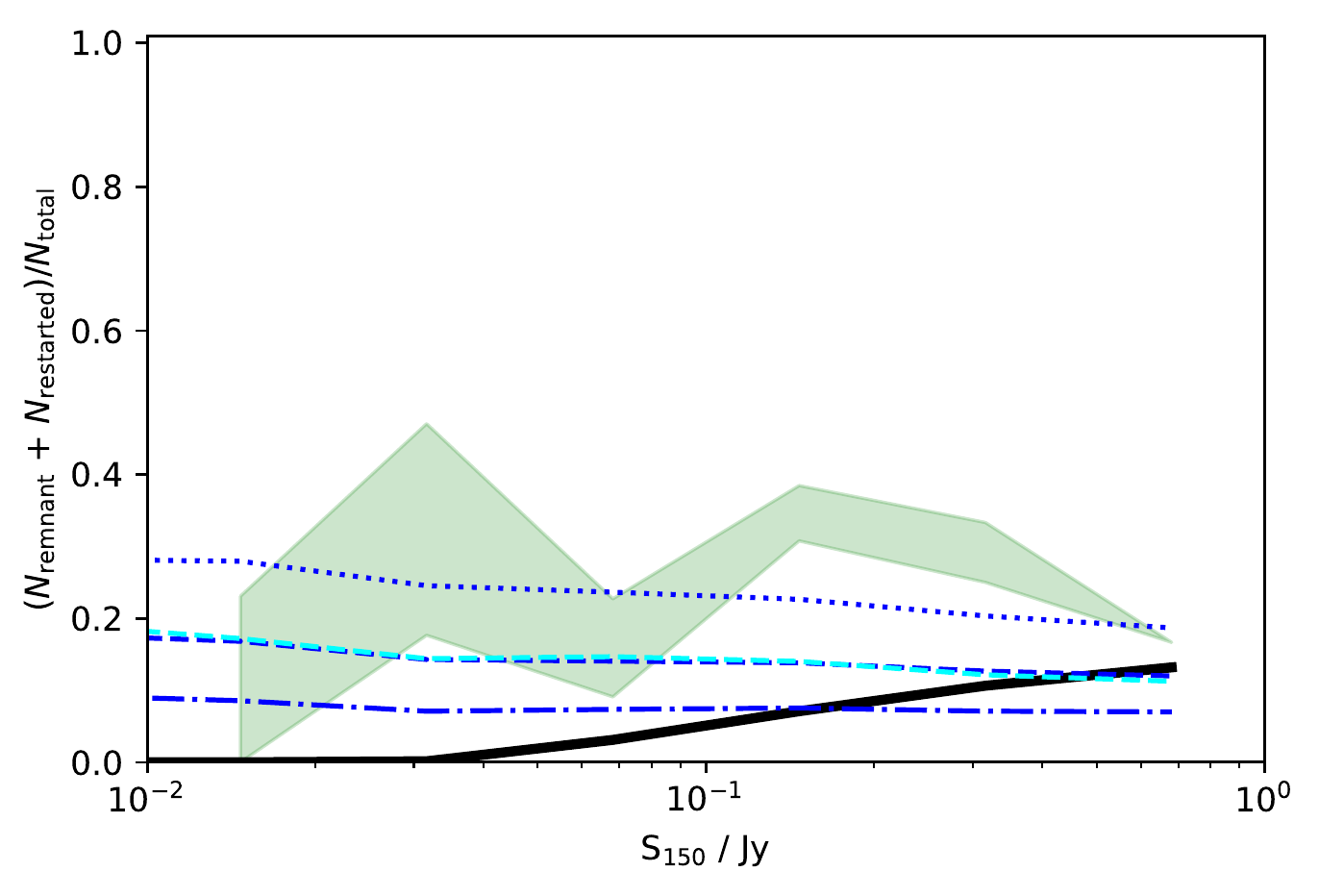}
\includegraphics[width=0.4\textwidth]{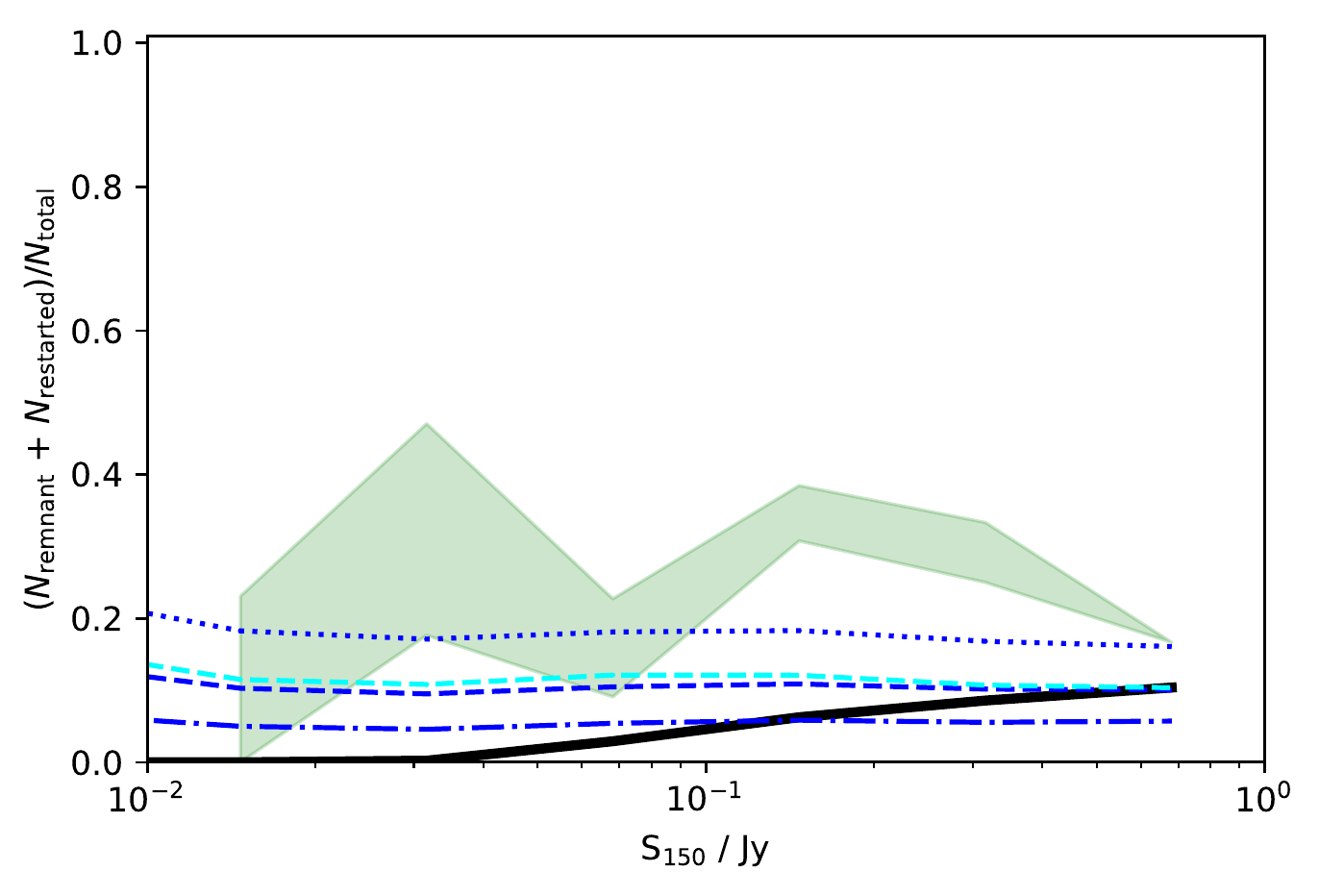}
\includegraphics[width=0.4\textwidth]{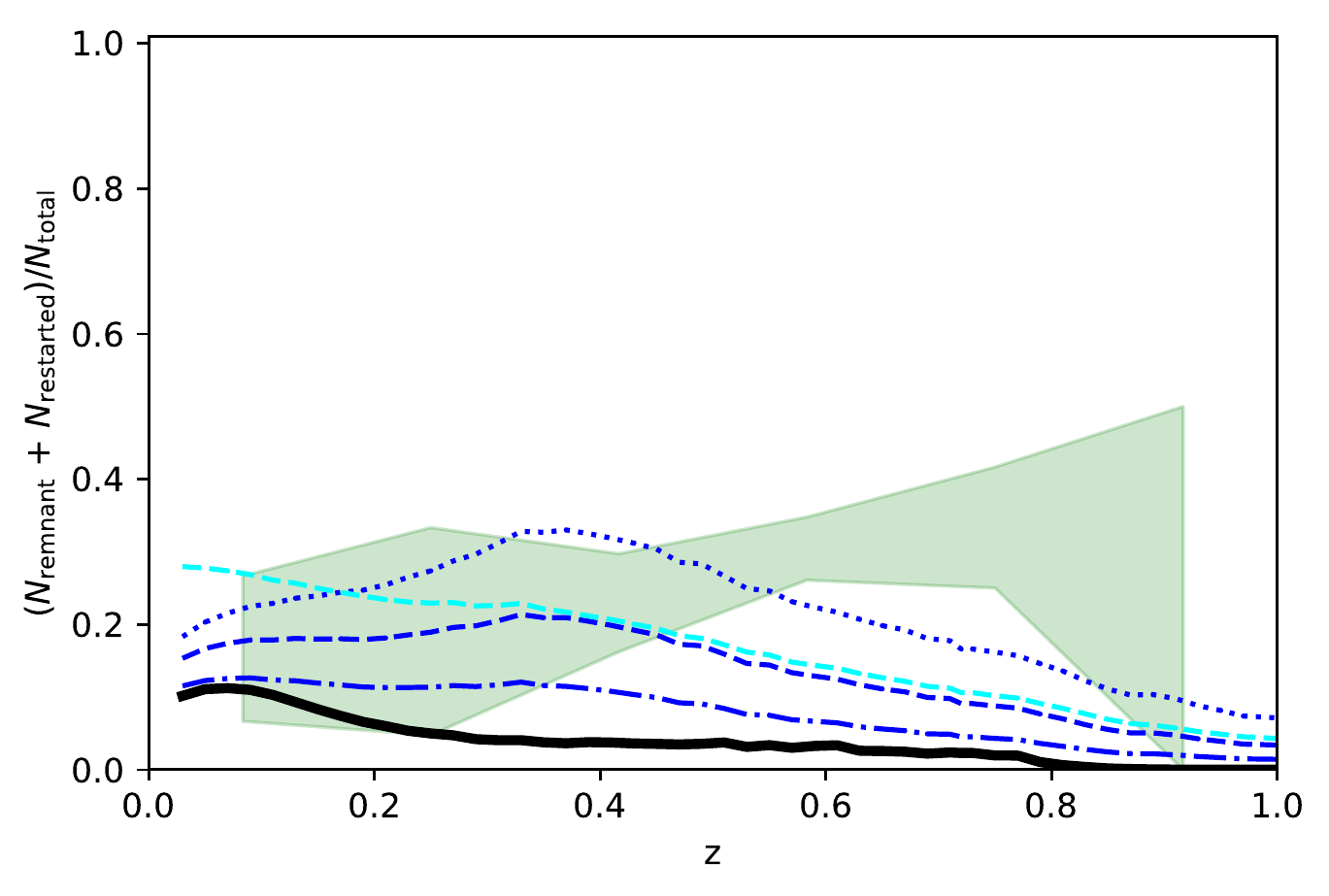}
\includegraphics[width=0.4\textwidth]{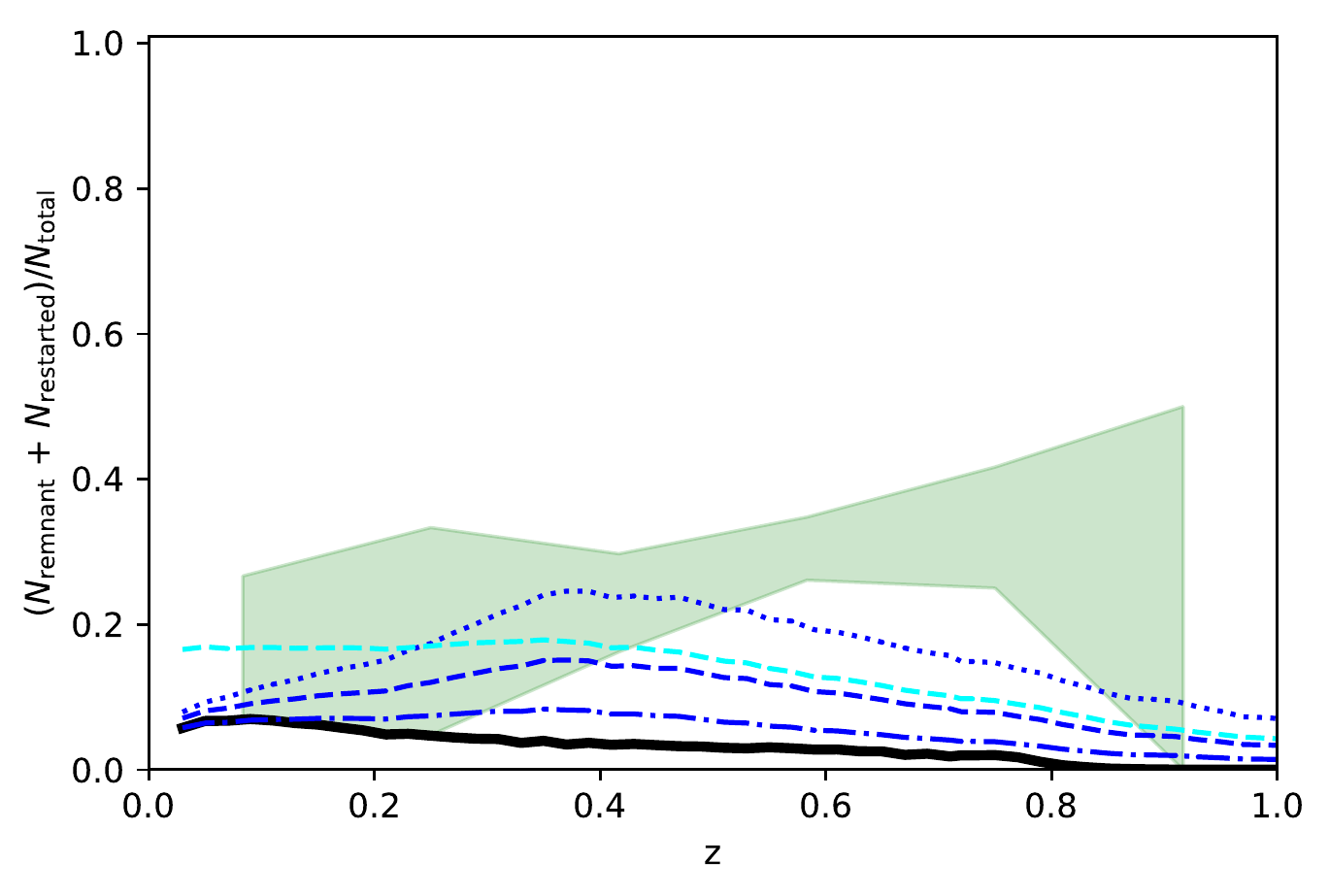}
\caption{Predicted remnant plus restarted fractions for FR-Is (left) and FR-IIs (right) for power-law age models. Models with a high fraction of short-lived sources are in better agreement with observations (green shaded regions) than constant age models.}
\label{fig:pred_remnantFrac_plAge}
\end{center}
\end{figure*}

We plot model predictions for power-law age distributions in Figure~\ref{fig:pred_remnantFrac_plAge}. These are clearly in better agreement with observations of candidate restarted sources than constant age models. In particular, a model with $p(\log t_{\rm on}) d \log t_{\rm on} \propto t_{\rm on}^{-1}$ and $p(\log Q_{\rm jet}) d \log Q_{\rm jet} \propto Q_{\rm jet}^{-1}$ is in excellent agreement with observations of both active (Figure~\ref{fig:pred_active_dist_best_plAge}) and remnant plus restarted (Figure~\ref{fig:pred_remnantFrac_plAge}) radio source populations.

\section{Discussion}
\label{sec:discussion}

\subsection{Comparison with previous work}

Our constant age models can be compared with related work by \citet{GodfreyEA17}, \citet{BrienzaEA17} and \citet{Hardcastle18}. The constant age models presented here predict that remnant lobes fade quickly below the detection limit once the radio jets switch off, similar to the findings of \citet{GodfreyEA17} and \citet{BrienzaEA17}. In line with predictions by \citet{GodfreyEA17,BrienzaEA17,Hardcastle18}, the expected remnant plus restarted fraction decreases with redshift. We note that the (calibrated with active source populations) FR-I and FR-II models in fact make very similar predictions, suggesting once again that uncertainties in adopted modelling parameters should not greatly influence our results.

The remnant plus restarted fractions predicted by our single age models are consistently lower than \citet{Hardcastle18}'s values of $\geq 0.3$. This comes directly from the constraint on the power spectrum of the jet kinetic power: to reproduce the observed properties of the active (progenitor) radio galaxy population we required $a \sim 1.0$; on the other hand, \citet{Hardcastle18} used a uniform distribution in $\log Q$, i.e. $a=0$. With such a power-law slope, our single age model predicts remnant fractions of up to 0.35 at $z=0$, however such a distribution in jet kinetic power is inconsistent with observations. [We note that this model of \citet{Hardcastle18} was presented for illustrative purposes only; detailed dynamical modelling of sources in the LOFAR HETDEX field by \citet{HardcastleEA19} suggests a value $a \sim 1.0$ is more appropriate for the bulk of the observed radio AGN population (see their Figure A5), consistent with the results presented here.]

Similarly, the high predicted remnant fractions by \citet{GodfreyEA17} and \citet{BrienzaEA17} are due to their model assumptions about the progenitor population, most importantly the short $t \sim 30-40$~Myr median active lifetime: as pointed out by these authors \citep[see Section 4.4.3 of][]{GodfreyEA17}, the predicted remnant fraction scales approximately inversely with this parameter, as sources which ``switch off'' while still young are detectable as remnants for a larger fraction of their total visible lifetime. We obtain similarly high predicted remnant plus restarted fractions in our short-lifetime models (e.g. $t_{\rm on}=100$ Myr, $a=0.4$). However, such models fail to reproduce the observed properties of the active source populations (Figure~\ref{fig:pred_active_dist_best}), selected from the same field as the remnant and restarted sources. 

\subsection{Constraining the jet duty cycle}

The key result of this work is that constant age models cannot simultaneously explain the observed properties of active, remnant and restarted populations, assuming the restarted population is non-negligible. By contrast, models in which the radio AGN population is dominated by sources with short duty cycles, are consistent with observations, as seen in Figure~\ref{fig:pred_remnantFrac_plAge}.

\subsubsection{Expectations from models}

Any successful constant age model must simultaneously satisfy two competing observational constraints. On the one hand, the radio sources must be sufficiently long-lived to give rise to the largest observed active radio galaxies (Figure~\ref{fig:pred_active_dist_best}). On the other hand, large, old radio remnants fade rapidly once the jets switch off (e.g. Figure~\ref{fig:example_tracks}), making it difficult to produce high observable remnant fractions. Conversely, models with many short-lived progenitors can produce high remnant fractions (Figure~\ref{fig:pred_remnantFrac_constAge}) but struggle to match the observed properties of their progenitor, active population (Figure~\ref{fig:pred_active_dist_best}) at the same time.

Power-law age models are a natural solution: the numerous short-lived sources produce a large population of remnant lobes which are visible for a long time (relative to the typical jet active lifetime); while the infrequent long-lived jet episodes -- the tail of the power-law distribution in age -- give rise to the relatively uncommon, large radio galaxies.

\subsubsection{Observational evidence for the dominance of short-lived sources}

Distributions of sizes and luminosities in complete radio source samples show an excess of compact sources over what would be expected from models in which all sources live to the same age (i.e. constant age models), particularly at lower radio luminosities \citep{ShabalaEA08,HardcastleEA19}. Further support for the dominance of short-lived jets comes from considering double-double radio sources. The compact (GPS/CSS) AGN phase is relatively short lived ($\leq$ several Myr), which is comparable to or shorter than the minimum time required for inner lobes of a double-double radio galaxy to catch up to the outer lobes\footnote{Expansion speed of the inner lobes is approximately limited to the sound speed of the relativistic lobe plasma from a previous jet episode, $\sim 0.3c$. Using a typical remnant size of 400~kpc implies lobe merging time $\geq 2$~Myr. Lower expansion speeds (e.g. if the cavity from a previous outburst has been refilled, and the jets must do work against the IGM/ICM) imply larger numbers of double-doubles.}. Hence, for a fixed duty cycle the restarted population should be dominated by double-doubles rather than restarted sources with compact cores. This is not seen in LOFAR observations \citep{JurlinEA20}.

Frequent re-triggering of radio jets is also suggested by two further recent observational studies. \citet{SabaterEA19} found that low-level jet activity is ubiquitous in massive galaxies; while \citet{BruniEA19} reported that the majority of high-power Giant Radio Galaxies show evidence of repeated jet activity.

\subsubsection{Jet triggering mechanisms}

Dynamical modeling of LOFAR radio source populations in the HETDEX field by \citet{HardcastleEA19} found some tentative evidence for a higher fraction of short-lived sources at low luminosities, potentially reflecting different jet triggering mechanisms for different radio source populations, as previously suggested by numerous authors \citep[e.g.][]{PimbbletEA13,KavirajEA15,MarshallEA18,KrauseEA19}. 

Why are most radio sources short-lived? A clue may lie in simulations of black hole -- galaxy co-evolution, which consistently predict that black hole accretion rates vary with time, and are regulated by the feedback (either mechanical or radiative) from the AGN. The power spectrum predicted in such feedback-regulated scenarios approaches pink noise \citep{NovakEA11,GaborBournaud13,GaspariEA17}, apparently consistent with the best-fitting power law age distribution derived in this work, $p(\log t_{\rm on}) d \log t_{\rm on} \propto t_{\rm on}^{-1}$.

It is tempting to suggest that low-power jets, which dominate complete samples, are more likely to be affected by the feedback-regulated gas cooling cycle through jet mass-loading and subsequent disruption \citep{Bicknell95,LaingBridle02,CrostonHardcastle14}. The cooling-regulated disruption of the jet, perhaps mediated by a mechanism similar to the Chaotic Cold Accretion proposed by \citet{GaspariEA17}, would then naturally lead to shorter duty cycles in these objects, and hence a higher fraction of restarted sources\footnote{Stronger entrainment in low-power sources will also result in their jet kinetic powers being systematically underestimated \citep{GodfreyShabala13,HardcastleEA19}, and hence feedback from these objects may be more important than current energetics estimates \citep[e.g.][]{TurnerShabala15,HardcastleEA19} suggest.}.

One potential caveat is that large (old) sources must have higher luminosities to exceed the surface brightness detection limit, and the apparent difference between high and low-luminosity populations may simply be a selection effect; environment \citep[e.g.][]{HardcastleKrause13,ShabalaEA17,Shabala18,KrauseEA19b} would be a further complicating factor. We defer a more detailed analysis to future work.

As discussed in \citet{JurlinEA20}, larger samples, more sensitive observations and broader frequency coverage are needed to identify with confidence compact, restarted jets in LOFAR data. Figures~\ref{fig:pred_remnantFrac_constAge} and \ref{fig:pred_remnantFrac_plAge} show that constant and power-law age models predict different distributions of remnant plus restarted fraction as a function of flux density, source size and redshift. Combining robust source classifications with dynamical models holds much promise for yielding deeper insights into the radio source duty cycles, and ultimately mechanisms responsible for the modulation of jet activity.

\section{Conclusions}
\label{sec:conclusions}

We have used dynamical radio source models to study the radio jet duty cycle in the Lockman Hole. Unlike previous work, we use observations of active radio galaxy populations to constrain the progenitors of radio remnants in our models. For our sample of moderately powerful radio sources we find the following results.

\begin{itemize}

\item{Active radio galaxy populations are equally well fitted by two different sets of models: (i) models in which all radio jets have a maximum lifetime ($\sim 300$ Myr using our assumed jet and environment parameters); and (ii) models with a distribution of source ages, $p(t_{on}) \propto t^{-1}$. For both sets of models, we require a power-law distribution of jet powers, $p(Q) \propto Q^{-1}$. FR-I and FR-II models make very similar predictions, due to the competing effects of particle content (more radiating particles for the same jet kinetic power in FR-IIs) and environment (lower external pressure and hence radio luminosity in FR-IIs).}

\item{Degeneracy between constant age and power-law age models can be broken by observations of remnant and restarted sources. Constant age models predict a short-lived detectable remnant phase. All models which match the observed properties of the progenitor population predict remnant plus restarted fractions $\lesssim 5$ percent. Predicted remnant / restarted fractions show a strong dependence on redshift, flux density and angular size.}

\item{Power-law age models predict much higher remnant / restarted fractions than constant age models. The predicted remnant / restarted fraction in power-law age models does not depend strongly on observables.}

%{\color{cyan}
%\item{The remnants are not expected to be ultra-steep. We find a median offset of $\alpha_{150}^{1400}=0.2-0.3$ between the spectral index of active and remnant populations. Once the jets switch off, remnants fade faster than their spectra steepen.}
%}

\item{ A high ($>10$ percent) fraction of genuine re-started sources would imply an appreciable fraction of short-lived and/or low power sources, qualitatively consistent with expectations from simulations of feedback-regulated black hole accretion.}

\end{itemize}

Model predictions of remnant fraction as a function of redshift, flux density and source size (as in Figure~\ref{fig:pred_remnantFrac_plAge}) provide a theoretical reference for ongoing, sensitive searches for low-luminosity remnant AGN with LOFAR and other telescopes. The combination of environment-sensitive radio source models and multi-wavelength data should constrain the physical properties of the radio jet populations, and ultimately quantify the role these objects play in galaxy evolution.

\section*{Acknowledgements}

We thank the anonymous referee for a constructive and prompt report, which helped improve the manuscript.

S.S. thanks the Australian Government for an Endeavour Fellowship 6719\_2018, and the Centre for Astrophysics Research at the University of Hertfordshire for their hospitality.

M.B. acknowledges support from the ERC-Stg DRANOEL, no 714245 and from INAF under PRIN SKA/CTA FORECaST.

LOFAR, the Low Frequency Array designed and constructed by ASTRON, has facilities in several countries, which are owned by various parties (each with their own funding sources), and that are collectively operated by the International LOFAR Telescope (ILT) foundation under a joint scientific policy.

The research leading to these results has received funding from the European Research Council under the European Union's Seventh Framework Programme (FP/2007-2013)/ERC Advanced Grant RADIOLIFE-320745. 

\bibliography{lofar_remnants.bib} % if your bibtex file is called example.bib
\bibliographystyle{mnras}

\end{document}